\documentclass{article}
\usepackage{slashed,verbatim}
\usepackage[numbers,sort&compress]{natbib}
\usepackage{amsmath}

\newcommand{\Rmnum}[1]{\expandafter\@slowromancap\romannumeral #1@}
\makeatother
\usepackage{amssymb}
\usepackage{float}
\usepackage{comment}
\usepackage{appendix}
\usepackage{tikz-feynman}
\usepackage{graphicx}
\usepackage{subcaption}
\usepackage{dcolumn}
\usepackage{bm}
\usepackage[colorlinks=true,linktocpage=true,
linkcolor=blue,citecolor=blue]{hyperref}
\usepackage[a4paper]{geometry}
\newcommand{\nd}{\noindent}
\newcommand{\be}{\begin{eqnarray}}
	\newcommand{\ee}{\end{eqnarray}}

\begin{document}
	\title{\bf{Momentum transport of chiral modes in a thermal QCD 
			medium}}
	% with coupling running with $T$- and $B$-dependent quark masses }}
	\author{Pushpa Panday\footnote{pushpa@ph.iitr.ac.in}~~and~~Binoy Krishna
Patra\footnote{binoy@ph.iitr.ac.in}\vspace{0.1in}\\
Department of Physics,\\
Indian Institute of Technology Roorkee, Roorkee 247667, India}
%\date{}
\maketitle
	
	\begin{abstract}
		We have investigated the effect of a weak magnetic field on momentum 
		transport in a thermal QCD medium at finite quark chemical potential 
		using a semiclassical kinetic theory.
		In the presence of a magnetic field, the momentum transport 
		coefficients acquire a tensor structure, reflected by the five shear 
		$\left(\eta_0, \eta_1, \eta_2, 
		\eta_3\right.$ and $\left.\eta_4\right)$ and two bulk viscous 
		components $\left(\zeta_0\right.$ and $\left. \zeta_1\right)$.
		The weak magnetic field removes the  degeneracy in the effective mass
		of flavours, leading to different masses for the left-handed (L) and 
		right-handed (R) chiral modes of quarks. The coefficients, $\eta_0, \eta_1$ and 
		$\eta_3$ decrease with magnetic field in L mode and increase in  
		$\mathrm{R}$ mode, whereas, $\eta_2$ and $\eta_4$ 
		increase in both L and $\mathrm{R}$ mode. The bulk viscous 
		coefficients, $\zeta_0$ and $\zeta_1$ increase with magnetic field for 
		both L and R mode.
		The shear and bulk viscous coefficients positively amplify with baryon 
		asymmetry. 
		
		\begin{comment}
			The specific shear viscosities, $\eta_0/s, \eta_1/s$ 
			and $\eta_3/s$
			exhibit the same trend with $B$ as their corresponding shear viscosities in 
			both chiral modes. However, $\eta_2/s$ and $\eta_4/s$ 
			follow the same trend as 
			$\eta_2$ and $\eta_4$ for L mode, and the opposite trend for R mode.
			Similarly, the specific bulk viscosities, $\zeta_0/s$ and $\zeta_1/s$, also 
			exhibit the same trend as $\zeta_0$ and $\zeta_1$, respectively, with magnetic field for both modes.
			The magnitude of L mode Reynolds number associated with $\eta_0, 
			\eta_1$ and $\eta_2$ is found to be greater than that of R 
			mode.
		\end{comment}
	\end{abstract}

	%\tableofcontents
	
	\section{Introduction}
	Ultrarelativistic heavy ion collisions at experimental facilities 
	such as Relativistic Heavy Ion Collider (RHIC) at Brookhaven National 
	Laboratory, Super Proton Synchrotron (SPS), Large Hadron Collider 
	(LHC) at European Organization for Nuclear Research, aim to produce the 
	new state of nuclear matter at extremely high temperature and/or 
	density. The composite states, {\em{hadrons}}, lose their identity and 
	dissolve into a soup of their constituents$-$ quarks and gluons known as 
	quark-gluon plasma (QGP). A very strong magnetic field ($eB$) is 
	produced during the initial stage of non-central heavy-ion collisions, 
	whose magnitude is of the order of $|eB| = 0.1 m_{\pi}^2$ ({$m_{\pi}\sim$ 0.135 GeV, pion mass}) for SPS 
	energy, $|eB|=m_{\pi}^2$ for RHIC and $|eB| = 15 m_{\pi}^2$ for LHC 
	\cite{kharzeev,skokov,toneev}. The possibility of existence of 
	such a strong magnetic field has motivated the theoreticians to study 
	the QGP under the intense field. The various novel phenomena such as 
	chiral magnetic effect \cite{kharzeev,fukushima}, magnetic and inverse 
	magnetic catalysis \cite{gusynin,gusy}, axial magnetic effect 
	\cite{braguta, maxim}, chiral vortical effect \cite{son,liao}, 
	dilepton production rate \cite{tuchin}, and various thermodynamical 
	\cite{suba_rath,bithika} and transport coefficients 
	\cite{suba,manu,shiyong,hattori,chen,nam,dey,panday} etc. have 
	been studied in the presence of a magnetic field.
	
	The magnetic field produced was considered to have been strong 
	for a very short span of time,  $t\sim 0.2 $ $\mathrm{fm/c}$ for RHIC 
	energies \cite{kharzeev,asakawa}, but it was then brought out 
	that its lifespan is prolonged by the finite electrical conductivity of the medium 
	\cite{kirill,killi}. The predictions of 
	charged hadron elliptic flow
	from RHIC \cite{paul} and their theoretical explanations using 
	dissipative hydrodynamics \cite{abelev} have provided the 
	experimental evidence of existence
	of the transport processes in the QGP. These
	transport coefficients are not directly measurable 
	experimentally, rather serve as input parameters in the theoretical 
	modeling of experimental observables such as directed flow,
	elliptic flow etc.
	%The transport coefficients are significant physical parameters that characterise the features of QGP and describe the nature of interactions between quarks and gluons. Since, it is widely known that the QGP formed in high-energy heavy-ion collisions behaves like a fluid \cite{raimond}, therefore the evolution of QGP following heavy-ion collisions can be explored using relativistic
	%hydrodynamics, which uses transport coefficients as input
	%parameters.
	\par
	One of the most important transport coefficients in the 
	hydrodynamical description is shear viscosity ($\eta$) that
	governs the rate of momentum transfer in a presence of
	inhomogeneity of fluid velocity, while the bulk viscosity 
	($\zeta$) describes the change of local pressure when the fluid
	element is either expanding or contracting. The relativistic 
	hydrodynamics played one of the most important roles to extract 
	the value of shear viscosity to entropy density ratio ($\eta/s$)
	from the available experimental data 
	\cite{shen,ulrike,paul,song,heinz,victor,denicol,gale},
	%It has also been supported by several calculations based on the viscous hydrodynamics \cite{bass} or AdS/QCD models \cite{policastro,buchel} implying that QGP is strongly coupled system \cite{arsene}.
	and the ratio is found to be below or close to the 
	Kovtun-Son-Starinets 
	(KSS) bound \cite{PhysRevLett.94.111601}, $\eta/s>\frac{1}{4\pi}$.
	On the other hand at the classical level, the bulk viscosity 
	vanishes for the conformal fluid and massless QGP but quantum effects 
	break the conformal symmetry and it
	acquires non-zero value even for the massless QGP found in the 
	lattice studies of SU(3) gauge
	theory \cite{meyer}. In several other studies, the value of 
	$\eta/s$ 
	was found to be minimum \cite{csernai,lacey} near the phase 
	transition 
	whereas for bulk viscosity was found to be maximum 
	\cite{karsch,paech}. 
	Taking into account all these studies, shear and bulk viscosities 
	for a 
	hot QCD medium  have been studied using various approaches 
	\textit{viz.} kinetic theory \cite{daniel,sasaki}, perturbative 
	QCD 
	\cite{arnold,peter,hidaka}, Kubo formalism \cite{snigdha}.\par
	The transport coefficients such as shear viscosity, bulk 
	viscosity and 
	electrical conductivity are taken as an input to dynamical model 
	such 
	as relativistic magnetohydrodynamics. Hence, it is important to 
	calculate these transport coefficients in the presence of a 
	magnetic 
	field. In the present work, we have studied the shear and bulk 
	viscosity in the presence of a weak magnetic field at finite 
	quark 
	chemical potential using kinetic theory. 
	%The relativistic dispersion relation of fermion of mass ($m$) in a uniform magnetic field (${\bf{B}} = B\hat{z}$) is given as
	%\begin{equation*}
	%E_n^2 = p_z^2 + m^2 + 2nqB,
	%\end{equation*}
	%where, $n= 0,1,2\dots$ denotes the Landau levels. The probability of fermions getting thermally excited to higher Landau levels 
	%is exponentially suppressed as $\exp\left(\frac{-\sqrt{qB}}{T}\right)$ \cite{k_fuku}. In strong magnetic field limit ($\sqrt{qB}>>T$), the fermions occupy only the lowest Landau level ($n=0$) and hence known as lowest Landau level approximation. 
	%In weak magnetic field limit ($\sqrt{qB}<<T$), temperature is the dominant scale. Hence, thermal energy is much larger than the energy level spacing ($\sim\sqrt{qB}$), which forces the fermions to occupy the higher Landau levels.
	The viscous dissipative tensor gets decompose into seven 
	components resulting from the breaking of rotational invariance in the 
	presence of a magnetic field. In the presence of a strong magnetic 
	field (${q_fB>>T^2,m_{f0}^2}$) with $T$, $q_f$ and $m_{f0}$ being temperature of the system, absolute electric charge and current quark mass of quark of $f$-th flavor respectively, the motion of charged particle is 
	restricted to the $1+1-$ dimensional Landau level dynamics. 
	Hence, only the longitudinal component contributes to the viscous 
	coefficients 
	\cite{hattori,suba,kurian,salman}. 
	%The momentum transport coefficients have been explored in the presence of a strong magnetic field ($q_fB>>T^2,m_f^2$) in different frameworks, \textit{namely}, Kubo formalism \cite{hattori}, relativistic kinetic theory \cite{suba}, effective fugacity model \cite{kurian}. Due to the $1+1-$ dimensional Landau level dynamics in strong magnetic field, only the longitudinal component contributes to the viscous coefficients. 
	Furthermore, explorations have been made in the presence of an 
	arbitrary 
	magnetic field in various models such as hadron resonance gas 
	model 
	\cite{ashutosh,arpan}, perturbative QCD \cite{shiyong}, kinetic 
	theory 
	\cite{Denicol_2018,PhysRevD.99.056017,Hattori_2022,shubha,jayanta}, linear sigma model \cite{ritesh}, in which 
	unlike the case of strong magnetic field, transverse and Hall 
	components also make a finite contribution to the viscous 
	coefficients.
	We have explored the effect of weak magnetic field 
	($T^2>q_fB>m_{f0}^2$)
	and baryon asymmetry ($\mu\neq 0$) on the momentum transport 
	coefficients using kinetic theory via the quasiparticle 
	description of 
	partons. Here, we have employed the general framework of projection tensors and written it in terms of hydrodynamical and magnetic degrees of freedom \cite{hess}, to calculate the viscous coefficients. In 
	Ref. \cite{shubha}, authors have incorporated the pure thermal mass to 
	incorporate the interaction, while we have used the thermally 
	generated mass with magnetic field correction. The dispersion 
	relation 
	of quasiparticles in weak magnetic field gives four collective 
	modes, 
	two from left-handed and two from right-handed modes. Various 
	properties of dispersion relations have been discussed in 
	Refs. \cite{aritra,weldon}. The degeneracy in left-handed (L) and right-handed (R)
	chiral 
	modes of quarks is lifted up due to their different masses, which is 
	in 
	contrast to the strong magnetic field case, where these modes are 
	degenerate. We have taken into account the medium-generated mass for both modes separately to estimate the momentum transport coefficients. We have further extended the discussion to study the 
	specific shear viscosity, specific bulk viscosity and Reynolds number.\par
	The present work is organized as follows: In Sec. 
	\ref{quasiparticle}, 
	we have discussed the quasiparticle model of partons and hence 
	evaluated the medium generated mass. In Sec. \ref{transport}, we 
	have 
	used this mass as an input parameter to calculate the momentum 
	transport coefficients in the presence of a weak magnetic field 
	using 
	kinetic theory under relaxation time approximation (RTA) for both 
	modes. In Sec. \ref{results}, we have discussed the result of 
	these 
	transport coefficients, \textit{namely}, shear and bulk 
	viscosities at 
	different values of magnetic field and quark chemical potential. 
	Further, we have discussed some applications of these viscosities 
	in 
	terms of specific shear and bulk viscosities, and Reynolds number in 
	Sec. \ref{applications}. Finally, we have summarized 
	the work in Sec. \ref{conc}.
	\section{Quasiparticle Model for Partons}\label{quasiparticle}
	The quasiparticle description of quarks and gluons encodes the 
	interaction among partons in the form of medium-generated masses. 
	There 
	exists different models  for quasiparticle description such as 
	Nambu-Jona-Lasinio (NJL) model \cite{nambu,jona,hatsuda,costa}, 
	Polyakov NJL model \cite{fuku,ghosh,abuki}, Gribov-Zwanziger 
	quantization \cite{nan,flork}, linear sigma model 
	\cite{chakraborty,ritesh} etc. However, we have adopted the 
	perturbative 
	thermal QCD approach in which the medium generated masses for
	quarks and gluons are obtained from the poles of dressed
	propagators calculated by their respective self-energies. In pure 
	thermal medium at finite 
	quark chemical potential ($\mu$), the thermally
	generated mass for quarks and gluons are obtained
	as \cite{bellac}
	\begin{align}\label{mth}
		& m_{th}^2 = \frac{1}{8}g^2C_F\Big(T^2 + \frac{\mu^2}{\pi^2}\Big),\\
		& m_g^2 = \frac{1}{6}g^2T^2\left(C_A + \frac{1}{2}N_f\right),
	\end{align}
	respectively. $C_F = \left(N_c^2 - 1\right)/2N_C = \frac{4}{3}$ 
	for $N_C =3$, $C_A (C_A=3)$ is the group factor, $N_f$ is the 
	number of flavor, $g$ is the QCD coupling constant 
	with $g^2 = 4\pi\alpha_s$, where $\alpha_s$ is the
	one-loop running coupling constant, which runs with 
	temperature as \cite{ayala}
	\begin{equation}
		\alpha_s(\Lambda^2) = \frac{1}{b_1\ln\Big(\frac{\Lambda^2}{\Lambda_{\overline{MS}}^2}\Big)},
	\end{equation}
	where $b_1 = \left(11 N_c - 2 N_f\right)/12\pi$ and 
	$\Lambda_{\overline{MS}} =$ 0.176 GeV. The renormalization 
	scale for quarks and gluons is chosen to 
	be $\Lambda_q$ = $2\pi\sqrt{T^2 + \mu^2/\pi^2}$ 
	and $\Lambda_g$ = $2\pi T$ respectively. The effect of weak magnetic field is incorporated through the medium generated mass of quarks in a weakly magnetized thermal medium, which is determined from the poles of dressed quark propagator.  The gluons remains unaffected in the presence of a magnetic field, hence there will be no magnetic field contributing term in their medium dependent mass.
	%The thermal mass of fermions 
	%at low ($p<<m_{th}$) and high momentum limit ($p>>m_{th}$) is obtained 
	%to be of the same order, $m_{th}\sim gT$, which can be observed from 
	%the dispersion relation of fermions in pure thermal medium as 
	%\cite{bellac,blaziot} 
	%\begin{align}
	%\omega_+(p) = m_{th} + \frac{p}{3}; \qquad  p<<m_{th}\\
	%\omega_+(p) = p + \frac{m^2_{th}}{p}; \qquad p>>m_{th}.
	%\end{align}
	
	\begin{comment}
		\nd The effective quark mass for $f$th 
		flavor can be written in terms of current quark mass ($m_{f0}$) 
		and 
		thermal mass ($m_{f,th}$) as \cite{bannur}
		\begin{equation}
			m^2_f = m_{f0}^2 + \sqrt{2}m_{f0}m_{f,th}+m_{f,th}^2.
		\end{equation}
	\end{comment}
	
	\begin{comment}
		The effective quark 
		mass in the presence of a magnetic field can be generalized to
		\begin{equation}\label{4}
			m^2_f = m_{f0}^2 + \sqrt{2}m_{f0}m_{fth,B}+m_{fth,B}^2,
		\end{equation}
		where $m_{fth,B}$ can be obtained by taking the static limit of 
		denominator of the dressed quark propagator in magnetic 
		field.
	\end{comment}
	
\begin{figure}[h]\centering
		\includegraphics[width=0.35\textwidth]{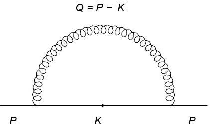}
		\caption{One-loop quark self energy in hot and magnetized medium}\label{self_quark}
	\end{figure}

	The inverse of the dressed quark propagator using 
	Schwinger-Dyson equation can be written as
	\begin{align}\label{inverse_prop}
		S^{*-1}(P) &= S^{-1}(P) - \Sigma(P)\nonumber\\
		&= \slashed{P} - \Sigma(P),
	\end{align}
	where $\slashed{P}=\gamma^{\mu}P_{\mu}$, $P$ is the momentum of external quark line, $S^{-1}(P)$ is bare inverse propagator and $\Sigma(P)$ is 
	the 
	quark self energy shown in Fig. \eqref{self_quark}. So, to calculate the effective quark propagator in presence of magnetic field at finite temperature we need to evaluate the quark self energy.
	\noindent The quark propagator in presence of background 
	magnetic field following the Schwinger formalism 
	can be written in terms of Laguerre 
	polynomial ($L_l(2\alpha)$) \cite{shwinger}
	\begin{equation}
		iS(K) = \sum_{l=0}^{\infty} \frac{-id_l(\alpha) D + d^{\prime}_l(\alpha)\bar{D}}{k_L^2+2l|q_fB|}+\frac{i\gamma\cdot k_{\perp}}{k_{\perp}^2},
	\end{equation}
	where $q_f$ is the absolute charge of $f$th flavor, 
	$l$ = 0, 1, 2, $\dots$ are the Landau levels, $||$ 
	and $\perp$ are the parallel and perpendicular 
	components of momentum respectively with respect 
	to direction of magnetic field, $\alpha=k_{\perp}^2/|q_fB|$, $k_L^2 = m_{f0}^2-k_{||}^2$ and $d_l(\alpha), d^{\prime}_l(\alpha), D, \bar{D}$ 
	are given as \cite{kang},
	
	\begin{align}
		&d_l(\alpha) = (-1)^l e^{-\alpha}C_l(2\alpha),\nonumber\\
		&d^{\prime}_l(\alpha) = \frac{\partial d_l}{\partial\alpha},\nonumber\\
		& D = (m_{f0}+\gamma \cdot k_{||})+ \gamma\cdot k_{\perp}\left(\frac{m_{f0}^2-k_{||}^2}{k_{\perp}^2}\right),\nonumber\\
		&\bar{D} = \gamma_1\gamma_2\left(m_{f0} + \gamma\cdot k_{||}\right),
	\end{align}
	with $C_l(2\alpha) = L_{l}(2\alpha) - L_{l-1}(2\alpha)$. In weak 
	field limit, the quark propagator can be reorganized 
	in power series of magnetic field $\left(q_fB\right)$ as,
	\begin{equation}\label{fermion_prop_magnetic}
		iS(K) = \frac{i\left(\slashed{K}+m_{f0}\right)}{K^2-m_{f0}^2} - \frac{\gamma_1\gamma_2\left(\gamma\cdot K_{||}+m_{f0}\right)}{\left(K^2-m_{f0}^2\right)^2}(q_fB),
	\end{equation}
	where first term in Eq.\eqref{fermion_prop_magnetic} is 
	the free fermion propagator and second term is the $\mathcal{O}(q_fB)$ correction to it. Neglecting the current quark mass 
	under the limit $\left(m_{f0}^2<q_fB<T^2\right)$ 
	in the numerator and using the following metric tensor in Eq.\eqref{fermion_prop_magnetic},
	\begin{align}
		g^{\mu\nu} = & g^{\mu\nu}_{||} + g^{\mu\nu}_{\bot};\nonumber\\
		g^{\mu\nu}_{\parallel} = \text{diag}(1,0,0,-1); &\quad g^{\mu\nu}_{\bot} = \text{diag}(0,-1,-1,0);\nonumber\\ 
		p^{\mu} = p^{\mu}_{\parallel} + p^{\mu}_{\bot}; &\quad p^{\mu}_{\parallel} = (p^0,0,0,p^3);\nonumber\\
		p^{\mu}_{\bot} = (0,p^1,p^2,0); &\quad \slashed{p} = \gamma^{\mu}p_{\mu}=\slashed{p}_{\parallel} + \slashed{p}_{\bot};\nonumber\\
		\slashed{p}_{\parallel} = \gamma^0 p_0 - \gamma^3 p^3; &\quad \slashed{p}_{\bot} = \gamma^1 p^1 + \gamma^2 p^2,
	\end{align}
	with
	\begin{equation}
		i\gamma_1\gamma_2\slashed{K}_{||} = -\gamma_5[(K\cdot b)\slashed{u}-(K\cdot u)\slashed{b}],
	\end{equation}
	we obtain the quark propagator in presence of magnetic field at finite temperature as
	\begin{equation}\label{quark_prop}
		iS(K) = \frac{i\slashed{K}}{K^2-m_{f0}^2} - \frac{i\gamma_5[(K\cdot b)\slashed{u}-(K\cdot u)\slashed{b}]}{(K^2-m_{f0}^2)^2}\big(q_f B\big),
	\end{equation}
	where $u^{\mu}= (1,0,0,0)$ denotes local rest frame of the heat bath. Introduction of a particular frame of reference breaks the Lorentz symmetry of the system. Similarly, $b^{\mu}=(0,0,0,1)$ denotes the preferred direction 
	of magnetic field in our system which then breaks the rotational 
	symmetry of the system. Using the quark propagator \eqref{quark_prop}, the one-loop quark self energy upto $\mathcal{O}(q_fB)$ 
	in hot and weakly magnetized medium can be written as
	\begin{align}\label{self_energy}
		\Sigma(P) =  g^2 C_F T\sum_n\int\frac{d^3k}{(2\pi)^3}\gamma_{\mu}\Bigg(\frac{\slashed{K}}{(K^2-m_{f0}^2)}
		-& \frac{\gamma_5[(K\cdot b)\slashed{u}-(K\cdot u)\slashed{b}]}{(K^2-m_{f0}^2)^2}(q_fB)\Bigg)\times&\nonumber\\
		\gamma^{\mu}\frac{1}{(P-K)^2},
	\end{align}
	where $T$ is the temperature of the system and $g^2 = 4\pi\alpha_s(\Lambda^2,|eB|)$ with $\alpha_{s}\left(\Lambda^2,|eB|\right)$ is given by
	\begin{equation}
		\alpha_s\left(\Lambda^2,|e B|\right)=\frac{\alpha_s\left(\Lambda^2\right)}{1+b_1 \alpha_s\left(\Lambda^2\right) \ln \left(\frac{\Lambda^2}{\Lambda^2+|e B|}\right)}.
	\end{equation}
	In Eq. \eqref{self_energy},	first term is the thermal medium contribution ($\Sigma_{0}$)  
	whereas second one is with magnetic field correction term ($\Sigma_{1}$).\par
	The general covariant structure of quark self energy 
	at finite temperature and magnetic field can 
	be written as \cite{aritra}
	\begin{equation}\label{52}
		\Sigma(P) = -\mathcal{A}\slashed{P}-\mathcal{B}\slashed{u}-\mathcal{C}\gamma_5\slashed{u}-\mathcal{D}\gamma_5\slashed{b},
	\end{equation} 
	where $\mathcal{A}, \mathcal{B}, \mathcal{C}, \mathcal{D}$ 
	are the structure functions. In the absecne of a magnetic field, $\mathcal{A}\neq 0, \mathcal{B}\neq 0$ but both $\mathcal{C},\mathcal{D}$ will vanish. In the presence of a pure magnetic field without any heat bath, $\mathcal{A}\neq0, \mathcal{B}=0$, and the structure functions $\mathcal{C}$ and $\mathcal{D}$ will depend upon the magnetic field, as we will see later. Using Eq.\eqref{self_energy} 
	and \eqref{52}, the general form of these structure 
	functions are obtained as 
	\begin{align}\label{A}
		& \mathcal{A}\left(p_0, p_{\perp}, p_z\right)= \frac{1}{4}\frac{\text{Tr}(\Sigma(P)\slashed{P})-(P\cdot u)\text{Tr}(\Sigma(P)\slashed{u})}{(P\cdot u)^2-P^2},\\ \label{B}
		& \mathcal{B}\left(p_0, p_{\perp}, p_z\right)= \frac{1}{4}\frac{\left(-P\cdot u\right)\text{Tr}(\Sigma(P)\slashed{P})+P^2\text{Tr}(\Sigma(P)\slashed{u})}{(P\cdot u)^2-P^2},\\ \label{C}
		& \mathcal{C}\left(p_0, p_{\perp}, p_z\right)= -\frac{1}{4}\text{Tr}(\gamma_5\Sigma(P)\slashed{u}),\\ \label{D}
		& \mathcal{D}\left(p_0, p_{\perp}, p_z\right)= \frac{1}{4}\text{Tr}(\gamma_5\Sigma(P)\slashed{b}).
	\end{align}
	These structure functions are found to depend upon 
	various Lorentz scalars defined by
	\begin{align}
		& p^0\equiv P^{\mu}u_{\mu}=\omega,\\
		& p^3\equiv P^{\mu}b_{\mu}= -p_z,\\
		& p_{\bot}\equiv\big[(P^{\mu}u_{\mu})^2 - (P^{\mu}b_{\mu})^2 - (P^{\mu}P_{\mu})^2\big]^{1/2},
	\end{align}
	where $\omega,p_{\bot},p_{z}$ are termed as Lorentz 
	invariant energy, transverse momentum and longitudinal 
	momentum respectively. The detailed calculation of 
	all these structure functions is shown in 
	\ref{append} and results are quoted here,
	\begin{align}\label{A_net}
		& \mathcal{A}(p_0,|{\bf{p}}|) = \frac{m_{th}^2}{|{\bf{p}}|^2}Q_1\Bigg(\frac{p_0}{|{\bf{p}}|}\Bigg),\\
		& \mathcal{B}(p_0,|{\bf{p}}|) =- \frac{m_{th}^2}{|{\bf{p}}|}\Bigg[\frac{p_0}{|{\bf{p}}|}Q_1\Big(\frac{p_0}{|{\bf{p}}|}\Big)-Q_0\Bigg(\frac{p_0}{|{\bf{p}}|}\Bigg)\Bigg],\\
		& \mathcal{C}(p_0,|{\bf{p}}|) = -4g^2C_FM^2\frac{p_z}{|{\bf{p}}|^2}Q_1\Bigg(\frac{p_0}{|{\bf{p}}|}\Bigg),\\
		& \mathcal{D}(p_0,|{\bf{p}}|) = -4g^2C_FM^2\frac{1}{|{\bf{p}}|}Q_0\Bigg(\frac{p_0}{|{\bf{p}}|}\Bigg),\label{D_net}
	\end{align}
	where $Q_0$ and $Q_1$ are Legendre functions of first
	and second kind respectively read as
	\begin{align}
		&Q_0(x) = \frac{1}{2}\ln\left(\frac{x+1}{x-1}\right),\\
		&Q_1(x) = \frac{x}{2}\ln\left(\frac{x+1}{x-1}\right)-1 = xQ_0(x)-1,
	\end{align}
	with magnetic mass obtained as \cite{aritra_bandho} 
	\begin{align}
		& M^2(T,\mu,m_{f0},q_fB) = \frac{|q_fB|}{16\pi^2}\left(\frac{\pi T}{2m_{f0}} -\text{ln} 2 + \frac{7\mu^2\zeta(3)}{8\pi^2T^2}\right),
	\end{align}
	where $\zeta$ is the Riemann zeta function. 
	The general 
	covariant structure of quark self energy Eq.(\ref{52})
	can be recast in terms of 
	left handed ($P_L =(\mathbb{I}-\gamma_5)/2$) 
	and right handed ($P_R =(\mathbb{I}+\gamma_5)/2$) 
	chiral projection operators as
	\begin{equation}\label{63}
		\Sigma(P) = -P_R\slashed{A'}P_L - P_L\slashed{B'}P_R,
	\end{equation}
	with $\slashed{A'}$ and $\slashed{B'}$ defined as
	\begin{align}
		& \slashed{A'} = \mathcal{A}\slashed{P} + (\mathcal{B}+\mathcal{C})\slashed{u} + \mathcal{D}\slashed{b},\\
		& \slashed{B'} = \mathcal{A}\slashed{P} + (\mathcal{B}-\mathcal{C})\slashed{u} - \mathcal{D}\slashed{b}.
	\end{align}
	
	Using Eq.\eqref{inverse_prop} and (\ref{63}), inverse fermion propagator can be written as
	\begin{align}
		S^{*-1}(P) = \slashed{P} + P_R\left[\mathcal{A}\slashed{P} + \left(\mathcal{B}+\mathcal{C}\right)\slashed{u} + \mathcal{D}\slashed{b}\right]P_L + P_L\left[\mathcal{A}\slashed{P} + \left(\mathcal{B}-\mathcal{C}\right)\slashed{u} - \mathcal{D}\slashed{b}\right]P_R,
	\end{align}
	and using $P_{L,R}\gamma^{\mu} = \gamma^{\mu} P_{R,L}$ and $P_L \slashed{P} P_L = P_R \slashed{P} P_R = P_L P_R \slashed{P} = 0$, 
	we obtain
	\begin{equation}
		S^{*-1}(P) = P_R \slashed{L} P_L + P_L \slashed{R} P_R,
	\end{equation}
	where $\slashed{L}$ and $\slashed{R}$ are 
	\begin{align}
		&\slashed{L} = (1+\mathcal{A})\slashed{P} + (\mathcal{B}+\mathcal{C})\slashed{u} + \mathcal{D}\slashed{b},\\
		& \slashed{R} = (1+\mathcal{A})\slashed{P} + (\mathcal{B}-\mathcal{C})\slashed{u} - \mathcal{D}\slashed{b}.
	\end{align}
	Thus, we get the effective quark propagator as
	\begin{equation}
		S^*(P) = \frac{1}{2}\left[ P_L\frac{\slashed{L}}{L^2/2}P_R+P_R\frac{\slashed{R}}{R^2/2}P_L\right],
	\end{equation}
	where 
	\begin{align}
		& L^2 = (1+\mathcal{A})^2P^2 + 2(1+\mathcal{A})(\mathcal{B}+\mathcal{C})p_0-2\mathcal{D}(1+\mathcal{A})p_z+ (\mathcal{B}+\mathcal{C})^2-\mathcal{D}^2,\\
		& R^2 = (1+\mathcal{A})^2P^2 + 2(1+\mathcal{A})(\mathcal{B}-\mathcal{C})p_0+2\mathcal{D}(1+\mathcal{A})p_z+ (\mathcal{B}-\mathcal{C})^2-\mathcal{D}^2.
	\end{align}
	For pure thermal medium, $\mathcal{C}$ and $\mathcal{D}$ will vanish which leads to $L^2=R^2$ or $L^{\mu}=R^{\mu}$. Therefore, the effective quark propagator becomes
	\begin{align}
		S^*(P)&=\frac{1}{R^2}\left[P_L\slashed{R}P_R+P_R\slashed{R}P_L\right],\\
		&=\frac{(1+\mathcal{A})\slashed{P}+\mathcal{B}\slashed{u}}{D},
	\end{align}
	where, $D=(1+\mathcal{A})^2P^2+2(1+\mathcal{A})\mathcal{B}P.u+\mathcal{B}^2$. Hence, $S^*(P)$ contains no chiral term. On the other hand, in the presence of a magnetic field at finite temperature, $L^2\neq R^2$ (parity violation \cite{weldon}) due to the non-vanishing value of $\mathcal{C}$ and $\mathcal{D}$, and hence the effective quark propagator contains the terms associated with $\gamma_{5}$.\par Next, we take the static limit 
	($p_0 =0,|{\bf{p}}|\rightarrow 0$) of $L^2/2$ and $R^2/2$,
	after expanding the Legendre functions involved 
	in structure functions in power series of 
	$\frac{|\bf{p}|}{p_0}$. Considering upto $\mathcal{O}(g^2)$, 
	we obtain
	\begin{align}
		&\frac{L^2}{2}\arrowvert_{p_0= 0, {|\bf{p}|}\rightarrow 0} = m_{th}^2 + 4g^2 C_F M^2,\\
		&\frac{R^2}{2}|_{p_0= 0, {|\bf{p}|}\rightarrow 0} = m_{th}^2 - 4g^2 C_F M^2.
	\end{align}
	Thus, the thermal mass (squared) at finite 
	chemical potential in presence of weak magnetic 
	field is obtained as
	\begin{align}\label{qaurk_mass}
		&m_{L}^2 = m_{th}^2 + 4g^2 C_F M^2,\\
		&m_{R}^2 = m_{th}^2 - 4g^2 C_F M^2,
	\end{align}
	where $m_{th}^2$, Eq. \eqref{mth} is the thermal mass in the absence of magnetic field. In the absence of magnetic field, $M^2=0$, therefore, $m_{L}^2=m_{R}^2=m_{th}^2$. In the weak magnetic field limit ($T^2>q_fB>m_{f0}^2$), 
	the magnetic field contribution to the medium generated quark mass appears as a correction to the pure thermal mass, $m_{th}$. The fact that $L^2\neq R^2$ in the presence of magnetic field, %and this non-equality of $L^2$ and $R^2$ appears as 
	leads to the %splitting in 
	lifting of the degeneracy of mass of chiral modes of quarks. %in weak magnetic field. 
	%Whereas 
	In  the case of a strong magnetic field, however, the quasiparticle mass of the quark does not appear as a correction to $m_{th}$, rather,
	the dominant contribution comes from the magnetic field. %Therefore, 
	On taking the static limit ($p_0=0, p_z\rightarrow 0$) of effective quark propagator in strong magnetic field limit, the medium generated mass (squared) of left and right handed chiral modes of quarks in strong magnetic field comes out as
	\cite{revisit}
	\begin{equation}
		m_{L,R}^2 =4g^2C_F\frac{|q_fB|}{16\pi^2}\left(\frac{\pi T}{2m_{f0}}-\text{ln}2\right).
	\end{equation}
	Now, we will assess the momentum 
	transport coefficients in the presence of a weak magnetic field 
	at a finite chemical potential for both L and R modes separately.
	\section{Momentum Transport Coefficients}\label{transport}
	In this section, we will explore the shear and bulk viscosity for 
	strongly interacting matter in a weakly magnetized thermal 
	medium. The Boltzmann transport equation governs the evolution of 
	single particle distribution $f(x,p)$ associated with partons in 
	our system. Since a deconfined medium of quarks and gluons is a relativistic system, therefore 
	it allow us to use the relativistic Boltzmann transport equation 
	(RBTE) in the partonic system. The RBTE for relativistic particle 
	with a charge $q$ in the presence of an external electromagnetic 
	field can be written as 
	\cite{landau}
	\begin{equation}\label{1}
		p^{\mu}\partial_{\mu}f(x,p) + q F^{\mu\nu}p_{\nu}\frac{\partial f(x,p)}{\partial p^{\mu}} = C[f],
	\end{equation}
	where $f(x,p)$ is the slightly deviated distribution function 
	from equilibrium distribution function $f_{eq}$ with $f = f_{eq} 
	+ \delta f (\delta f<<f_{eq})$. $F^{\mu\nu}$ is the 
	electromagnetic field tensor 
	which in the absence of an electric field becomes, $F^{\mu\nu} = 
	-B b^{\mu\nu}$, where $b^{\mu\nu} = \epsilon^{\mu\nu\alpha\beta} 
	b_{\alpha}u_{\beta}$ with unit four vector, $b^{\mu} = 
	\frac{B^{\mu}}{B}$. The rate of change of distribution function 
	by means of collision is described by collision integral $C[f]$, 
	whose general form consists of absorption and emission terms in 
	phase space volume element. This leads to the nonlinear 
	integro-differential equation which is very cumbersome to solve. 
	Therefore, we simplify the equation by employing RTA \cite{witting}, in which 
	the external perturbation pushes the system slightly out of 
	equilibrium, from which it returns exponentially to equilibrium 
	with time scale $\tau$. Eq.\eqref{1} under 
	RTA takes the form as
	\begin{equation}\label{48}
		p^{\mu}\partial_{\mu}f(x,p) + q F^{\mu\nu}p_{\nu}\frac{\partial f(x,p)}{\partial p^{\mu}} = -\frac{p^{\mu}u_{\mu}}{\tau}\delta f,
	\end{equation}
	where first order correction is found to be
	\begin{align}
		\delta f_{1}=& -\frac{\tau}{u\cdot p}\left(p^{\mu}\partial_{\mu}+qF^{\mu\nu}p_{\nu}\frac{\partial}{\partial p^{\mu}}\right)f_{eq},\\\nonumber
		&=-\frac{\tau}{u\cdot p}p^{\mu}\partial_{\mu}f_{eq};\qquad qBb^{\mu\nu}p_{\nu}\frac{\partial f_{eq}}{\partial p^{\mu}}=0.
	\end{align}
	Since, first order correction does not take into account the effect of magnetic field therefore, we need to take a correction term in $qF^{\mu\nu}p_{\nu}\frac{\partial}{\partial p^{\mu}}$ to see the effect of magnetic field. For small deviation from equilibrium $(\delta f<<f_{eq})$, the 
	Eq.\eqref{48} can be written as
	\begin{equation}\label{49}
		p^{\mu}\partial_{\mu}f_{eq} = -\frac{p^{\mu}u_{\mu}}{\tau}\left[1-\frac{qB\tau b^{\mu\nu}p_{\nu}}{p^{\mu}u_{\mu}}\frac{\partial}{\partial p^{\mu}}\right]\delta f,
	\end{equation}
	$f_{eq}$ is the equilibrium distribution function, defined as
	\begin{equation}
		f_{eq}=\frac{1}{e^{(\sqrt{{\bf{p^2}}+m^2}-\mu)/T}+\alpha},
	\end{equation}
	where, $\varepsilon = 
	\sqrt{\textbf{p}^2+m^2}$ is the single particle energy and $\alpha=-1,+1,0$ is for boson, fermion and Boltzmann gas, respectively. \par
	The conserved net particle four-current ($N^{\mu} = nu^{\mu} 
	+ n^{\mu}$) and energy-momentum tensor 
	($T^{\mu\nu}=T^{\mu\nu(0)}+\Delta 
	T^{\mu\nu} = T^{\mu\nu(0)} -\Pi\Delta^{\mu\nu}+\pi^{\mu\nu}$), 
	with $u^{\mu}$ is defined in Landau frame, 
	can be expressed in terms of single particle phase-space distribution function as
	\begin{align}
		&N^{\mu} =\int dP~p^{\mu}(f-\bar{f}),\\\nonumber
		&T^{\mu\nu} = \int dP~p^{\mu}p^{\nu}(f+\bar{f}).
	\end{align}	
	Here, $dP= g~d^3p/\left[(2\pi)^3\sqrt{\bm{p}^2+m^2}\right]$, $g$ and $m$ being the degeneracy factor and particle rest mass, $p^{\mu}$ is the particle four-momentum. The derivative 
	($\partial_{\mu}$) can be split up covariantly into time and 
	space derivative: $\partial_{\mu}= u_{\mu}D + \nabla_{\mu}$, where
	$D = u^{\mu}\partial_{\mu}$ and $\nabla_{\mu} = 
	\Delta_{\mu\nu}\partial^{\nu} =\partial_{\mu} - u_{\mu}D 
	$,
	with $\Delta^{\mu \nu} = 
	g^{\mu\nu}-u^{\mu}u^{\nu}$. 
	Under this decomposition, the left-hand side derivative of 
	Eq.\eqref{49} can be expressed in terms of derivative of 
	thermodynamic parameters as
	\begin{align}\label{50}
		p^{\mu}\partial_{\mu}f_{eq} &= p^{\mu}\left(u_{\mu}D+\nabla_{\mu}\right)f_{eq}\\\nonumber
		&=\left(\frac{\partial f_{eq}}{\partial T}\right)\Big[(u\cdot p)DT + p^{\mu}\left(\nabla_{\mu}T\right)\Big]+\left(\frac{\partial f_{eq}}{\partial(\mu/T)}\right)\Big[(u\cdot p)D\left(\frac{\mu}{T}\right)+\\\nonumber &p^{\mu}\nabla_{\mu}\left(\frac{\mu}{T}\right)\Big]
		+\left(\frac{\partial f_{eq}}{\partial u^{\nu}}\right)\Big[\left(u\cdot p\right)Du^{\nu}+p^{\mu}\nabla_{\mu}(u^{\nu})\Big].
	\end{align}
	Since, the thermodynamic forces do not contain the time 
	derivative terms therefore the terms like $DT$ and 
	$D\left(\frac{\mu}{T}\right)$ 
	need to be expressed in terms of spatial derivative of 
	thermodynamic parameters. This can be achieved by making the use 
	of particle number conservation, 
	energy-momentum conservation and relativistic Gibbs-Duhem 
	relation.\par

	Therefore, Eq.\eqref{50} takes the form as
	\begin{align}\label{51}
		p^{\mu}\partial_{\mu}f_{eq} = \frac{f_{eq}\left(1-\alpha f_{eq}\right)}{T}\Bigg[&X\nabla_{\mu}u^{\mu}-p^{\mu}p^{\nu}\left(\nabla_{\mu}u_{\nu}-\frac{1}{3}\Delta_{\mu\nu}\nabla_{\sigma}u^{\sigma}\right)+\\\nonumber
		&\left(1-\frac{u\cdot p}{h}\right)Tp^{\mu}\nabla_{\mu}\left(\frac{\mu}{T}\right)\Bigg],
	\end{align}
	where $X=(u\cdot p)^2\left(\frac{4}{3}-\lambda'\right)-\frac{1}{3}m^2+\left(u\cdot p\right)\left[\left(\lambda''-1\right)h-\lambda'''T\right]$ and $h$ is 
	the 
	enthalpy per particle. The details of the calculation along with 
	expressions of $\lambda', \lambda''$ and $\lambda'''$ are given 
	in \ref{append_b}.
	
	The term $\delta f$ in Eq.\eqref{49} characterizes the deviation 
	of the distribution function from equilibrium distribution 
	function. $\delta 
	f$ can be constituted as linear combination of the
	thermodynamic forces ($Y_{\mu\nu}$) times appropriate tensorial 
	coefficients, resulting in a Lorentz scalar, $\delta f$ as,
	\begin{equation}
		\delta f=\mathrm{A} Y+\mathrm{B}^{\mu} Y_{\mu}+\mathrm{C}^{\mu \nu} Y_{\mu \nu}.
	\end{equation}
	By substituting the above form of $\delta f$ in Boltzmann 
	transport equation and comparing the coefficients of 
	thermodynamic forces, we obtain the unknown coefficients 
	$\mathrm{A}, \mathrm{B}^{\mu}$ and 
	$\mathrm{C}^{\mu\nu}$ in the expression of $\delta f$; and then 
	by incorporating the $\delta f$ in the thermodynamic flows, we 
	can obtain the transport coefficients. We are interested in the 
	computation of momentum transport coefficients, specifically, shear and 
	bulk viscosity, as discussed in the following subsection.
	\subsection{Shear Viscosity}
	The general form of $\delta f$ for shear viscosity in the 
	presence of a magnetic field can be expressed in terms of fourth-rank projection tensors as \cite{landau, ashutosh, arpan}
	\begin{align}\label{assump}
		\delta f = &\sum_{r=0}^{4}a_r A_{\mu\nu\alpha\beta}^{(r)}p^{\mu}p^{\nu}V^{\alpha\beta}\\\nonumber
		=& \Big[a_0P^{(0)}_{<\mu\nu>\alpha\beta} + a_1\left(P^{(1)}_{<\mu\nu>\alpha\beta}+P^{(-1)}_{<\mu\nu>\alpha\beta}\right)+ia_2\left(P^{(1)}_{<\mu\nu>\alpha\beta}-P^{(-1)}_{<\mu\nu>\alpha\beta}\right)+\\\nonumber
		&a_3\left(P^{(2)}_{<\mu\nu>\alpha\beta}+P^{(-2)}_{<\mu\nu>\alpha\beta}\right)+ia_4\left(P^{(2)}_{<\mu\nu>\alpha\beta}-P^{(-2)}_{<\mu\nu>\alpha\beta}\right)\Big]p^{\mu}p^{\nu}V^{\alpha\beta},
	\end{align}
	where $V^{\alpha\beta} = \frac{1}{2}\left(\frac{\partial 
		u^{\alpha}}{\partial x^{\beta}}+\frac{\partial 
		u^{\beta}}{\partial 
		x^{\alpha}}\right)$ and $P^{(r)}_{<\mu\nu>\alpha\beta} = 
	P^{(r)}_{\mu\nu\alpha\beta} +P^{(r)}_{\nu\mu\alpha\beta}$. In 
	general, 
	the fourth rank projection tensor is defined in terms of second 
	rank projection tensor as \cite{hess},
	\begin{equation}
		P_{\mu \nu, \mu^{\prime} \nu^{\prime}}^{(\mathrm{m})}=\sum_{\mathrm{m}_{1}=-1}^{1} \sum_{\mathrm{m}_{2}=-1}^{1} P_{\mu \mu^{\prime}}^{\left(\mathrm{m}_{1}\right)} P_{\nu \nu^{\prime}}^{\left(\mathrm{m}_{2}\right)} \delta\left(m, m_{1}+m_{2}\right),
	\end{equation}
	and second rank projection tensor is defined as,
	\begin{equation}
		\begin{aligned}
			P_{\mu \nu}^{0} &=b_{\mu} b_{\nu}, \\
			P_{\mu \nu}^{1} &=\frac{1}{2}\left(\Delta_{\mu \nu}-b_{\mu} b_{\nu}+i b_{\mu \nu}\right), \\
			P_{\mu \nu}^{-1} &=\frac{1}{2}\left(\Delta_{\mu \nu}-b_{\mu} b_{\nu}-i b_{\mu \nu}\right).
		\end{aligned}
	\end{equation}
	The properties of second rank projection tensor are as follows
	\begin{equation}
		\begin{gathered}
			P_{\mu \sigma}^{(\mathrm{m})} P_{\sigma \nu}^{\left(\mathrm{m}^{\prime}\right)}=\delta_{\mathrm{mm}^{\prime}} P_{\mu \nu}^{(\mathrm{m})}, \\
			\left(P_{\mu \nu}^{(\mathrm{m})}\right)^{\dagger}=P_{\mu \nu}^{(-\mathrm{m})}=P_{\nu \mu}^{(\mathrm{m})}, \\
			\sum_{\mathrm{m}=-1}^{1} P_{\mu \nu}^{(\mathrm{m})}=\delta_{\mu \nu}, \quad P_{\mu \mu}^{(\mathrm{m})}=1.
		\end{gathered}
	\end{equation}
	The left-hand side of Eq.\eqref{49} can be expressed in terms of 
	4-rank projection tensors 
	$P^{(r)}_{\langle\mu\nu\rangle\alpha\beta}$ as
	\begin{equation}\label{left_boltz}
		-\frac{p^{\mu}p^{\nu}}{2T}f_{eq}\left(1-\alpha f_{eq}\right)V^{\alpha\beta}\left(P^{(0)}_{\langle\mu\nu\rangle\alpha\beta}+P^{(1)}_{\langle\mu\nu\rangle\alpha\beta}+P^{(-1)}_{\langle\mu\nu\rangle\alpha\beta}+P^{(2)}_{\langle\mu\nu\rangle\alpha\beta}+P^{(-2)}_{\langle\mu\nu\rangle\alpha\beta}\right).
	\end{equation}
	Substituting $\delta f$ on right hand side of Eq.\eqref{49},
	\begin{equation}\label{right_boltz}
		-\left(\frac{u\cdot p}{\tau}\right)\left[1-\frac{qB\tau b^{\mu\nu}p_{\nu}}{u\cdot p}\frac{\partial}{\partial p^{\mu}}\right]\sum_{r=0}^{4}a_r A_{\alpha\beta\rho\sigma}^{(r)}p^{\alpha}p^{\beta}V^{\rho\sigma}
	\end{equation}
	and equating the coefficients of 
	$P^{(r)}_{\langle\mu\nu\rangle\alpha\beta}$ in Eq.\eqref{left_boltz} and \eqref{right_boltz}, we get the coefficients for a system quarks of multiple charge species as \cite{arpan, ashutosh,Tuchin:2011jw}
	\begin{align}
		&a_0 = \sum_f\frac{1}{2T}\frac{f_{eq,f}\left(1-f_{eq,f}\right)\tau_f}{(u\cdot p)},\\
		&a_1 = \sum_f \frac{\left(u\cdot p\right)f_{eq,f}\left(1-f_{eq,f}\right)\tau_f}{2T\left[\left(u\cdot p\right)^2+\left(q_fB\tau_f\right)^2\right]},\\
		&a_2 =\sum_f \frac{\left(q_fB\right)f_{eq,f}\left(1-f_{eq,f}\right)\tau_f^2}{2T\left[\left(u\cdot p\right)^2+\left(q_fB\tau_f\right)^2\right]},\\
		&a_3 =\sum_f \frac{\left(u\cdot p\right)f_{eq,f}\left(1-f_{eq,f}\right)\tau_f}{2T\left[\left(u\cdot p\right)^2+\left(2q_fB\tau_f\right)^2\right]},\\
		&a_4 =\sum_f \frac{\left(q_fB\right)f_{eq,f}\left(1-f_{eq,f}\right)\tau_f^2}{T\left[\left(u\cdot p\right)^2+\left(2q_fB\tau_f\right)^2\right]}.
	\end{align}
	where, $f$ stands for flavor. Here we have used  $f=$ up (u) and down (d). Similarly, the above coefficients for antiquarks can be obtained 
	using, 
	$\bar{f}_{eq}=\frac{1}{e^{(\sqrt{{\bf{p^2}}+m^2}+\mu)/T}+1}, 
	q_f\rightarrow q_{\bar{f}}$, $\tau_f\rightarrow\tau_{\bar{f}}$ 
	and 
	$\delta\bar{f} = \sum_{r=0}^{4}\bar{a}_r 
	A_{(r)\mu\nu\alpha\beta}p^{\mu}p^{\nu}V^{\alpha\beta}$.
	
	\begin{comment}
		\begin{align}
			&\bar{a}_0 = \sum_f\frac{1}{2T}\frac{\bar{f}_{eq,f}\left(1-\bar{f}_{eq,f}\right)\tau_{\bar{f}}}{(u.p)},\\
			&\bar{a}_1 = \sum_f \frac{\left(u.p\right)\bar{f}_{eq,f}\left(1-\bar{f}_{eq,f}\right)\tau_{\bar{f}}}{2T\left[\left(u.p\right)^2+\left(q_{\bar{f}}B\tau_{\bar{f}}\right)^2\right]},\\
			&\bar{a}_2 =\sum_f \frac{\left(q_{\bar{f}}B\right)\bar{f}_{eq,f}\left(1-\bar{f}_{eq,f}\right)\tau_{\bar{f}}^2}{2T\left[\left(u.p\right)^2+\left(q_{\bar{f}}B\tau_{\bar{f}}\right)^2\right]},\\
			&\bar{a}_3 =\sum_f \frac{\left(u.p\right)\bar{f}_{eq,f}\left(1-\bar{f}_{eq,f}\right)\tau_{\bar{f}}}{2T\left[\left(u.p\right)^2+\left(2q_{\bar{f}}B\tau_{\bar{f}}\right)^2\right]},\\
			&\bar{a}_4 =\sum_f \frac{\left(q_{\bar{f}}B\right)\bar{f}_{eq,f}\left(1-\bar{f}_{eq,f}\right)\tau_{\bar{f}}^2}{T\left[\left(u.p\right)^2+\left(2q_{\bar{f}}B\tau_{\bar{f}}\right)^2\right]},
		\end{align}	
	\end{comment}
	
	$\tau_{f(\bar{f})}$ is the relaxation time for quarks 
	(antiquarks)  
	expressed as follows \cite{kajantie} 
	\begin{equation}
		\tau_{f(\bar{f})}=\frac{1}{5.1 T \alpha_{s}^{2}\left(\Lambda^2,|eB|\right) \log \left(\frac{1}{\alpha_{s}\left(\Lambda^2,|eB|\right)}\right)\left(1+0.12\left(2 N_{f}+1\right)\right)}.
	\end{equation}
	It was argued in \cite{berrehrah} that finite parton mass has 
	little effect on scattering cross section and hence on relaxation 
	time. This leads to the qualitatively same result for massless and
	massive partons. The current light quark ($m_{u,d}$) masses are
	chosen to be 0.1 times the strange quark mass ($m_{s0}$) which
	is in compliance with chiral perturbation theory 
	\cite{cheng,schmidt}. 
	The parameters were adjusted to get the best fitted lattice
	data with $m_{s0} = 80$ MeV \cite{zhu}.\par The general form of 
	shear viscous tensor ($\pi^{\mu\nu}$) in terms of 4-rank 
	projection tensor can be written as
	\begin{eqnarray}\label{69}
		{\pi^{\mu\nu} = \eta^{\mu\nu\alpha\beta}\sigma_{\alpha\beta}},&\\\nonumber
		={\sum_{r=0}^{4} \eta_{(r)}A_{(r)}^{\mu\nu\alpha\beta}\sigma_{\alpha\beta}}.
	\end{eqnarray}
	where $\sigma^{\alpha \beta} = V^{\alpha \beta} - \theta \Delta^{\alpha \beta}/3$, $\theta=\nabla_{\mu}u^{\mu}$ is the scalar expansion. Employing the integral form of shear viscous tensor for quark and 
	antiquark
	\begin{equation}\label{70}
		\pi^{\mu\nu} =\Delta^{\mu\nu}_{\alpha\beta} \int dP~p^{\alpha}p^{\beta}\left(\delta f + \delta\bar{f}\right),
	\end{equation}
	and using Eq. \eqref{assump},
	we obtained the five shear viscosity components as follows:
	\begin{align}
		&\eta_0 =\sum_{f}\beta\tau_f~J^{(1)+}_{42},\\
		&\eta_1=\sum_{f}\beta\tau_f~Y^{(1)+}_{42},\\
		&\eta_2 = \sum_{f}\beta q_fB~\tau_f^2~Y_{42}^{(2)-},\\
		&\eta_3 = \sum_{f}\beta\tau_f~Z^{(1)+}_{42},\\
		&\eta_4 = \sum_{f}2\beta q_fB~\tau_f^2~Z_{42}^{(2)-}.
	\end{align}
	Here, $J_{42}^{(1)+}, Y_{42}^{(1)+}, Y_{42}^{(2)-}, Z_{42}^{(1)+}, Z_{42}^{(2)-}$ are thermodynamic integrals (defined in \ref{append_c}). One may compare the above coefficients qualitatively with Ref. \cite{shubha,ankit}. The difference arises due to the incorporation of quasiparticle mass in weak magnetic field limit. Depending upon the direction of magnetic field and current, we can have longitudinal, transverse and Hall effect. $\eta_0$ is the longitudinal component, 
	$\eta_1,\eta_3$ are the transverse components and $\eta_2, 
	\eta_4$ are the Hall components of shear viscosity. In the limit of vanishing magnetic field, $qB\rightarrow$0, $\eta_2$ and $\eta_4$ reduces to zero while $\eta_0$, $\eta_1$ and $\eta_3$ becomes equal as expected. We shall take into account the quasiparticle mass that 
	was shown to differ for L and R chiral modes of quarks. Now, we 
	will calculate the gluonic contribution to the shear viscosity. 
	Since, gluons do not interact with electromagnetic field 
	therefore there will be no magnetic field 
	dependent term and shear viscosity for gluons turns out to be
	\begin{equation}
		\eta_g = 2\beta\tau_gJ_{42}^{(1)},
	\end{equation}
	where, 
	$\tau_g$ is the relaxation time for gluons \cite{kajantie},
	\begin{equation}
		\tau_{g}=\frac{1}{22.5 T \alpha_{s}^{2}\left(\Lambda^2\right) \log \left(\frac{1}{\alpha_{s}\left(\Lambda^2\right)}\right)\left(1+0.06 N_{f}\right)}.
	\end{equation}
	\subsection{Bulk Viscosity}
	Now, we will discuss the computation of bulk viscosity in a 
	weakly magnetized thermal medium. There are three components of 
	bulk viscosity in the presence of a magnetic field. The general 
	form of $\delta f$ 
	associated to them is
	\begin{align}\label{assum}
		\delta f =&\sum_{r=1}^{3} c_{r} A_{(r)\mu\nu}\partial^{\mu}u^{\nu}\\\nonumber
		=&{\Big(c_1P^{\left(0\right)}_{\mu\nu} + c_2\left(P^{\left(1\right)}_{\mu\nu}+P^{\left(-1\right)}_{\mu\nu}  \right)+c_3\left(P^{\left(1\right)}_{\mu\nu}-P^{\left(-1\right)}_{\mu\nu}  \right) \Big)\partial^{\mu}u^{\nu}}.
	\end{align}
	Using the properties of second rank projetion tensor, the right-hand side of Eq.\eqref{49} becomes,
	\begin{equation}\label{81}
		{-\frac{u\cdot p}{\tau}\left\{c_1\left(b_{\mu}b_{\nu}\right)+c_2\left(\Delta_{\mu\nu}-b_{\mu}b_{\nu}\right)+ic_3b_{\mu\nu}\right\}\left(\partial^{\mu}u^{\nu}\right)},
	\end{equation}	
	and
	equating 
	the coefficients of $\Delta_{\mu\nu}\partial^{\mu}u^{\nu}, 
	b_{\mu}b_{\nu}\partial^{\mu}u^{\nu}$ and 
	$b_{\mu\nu}\partial^{\mu}u^{\nu}$ from Eq.\eqref{51} and 
	\eqref{81}, we 
	get \cite{ashutosh}
	\begin{align}
		{c_1} =& {c_2 = -\sum_f\frac{\tau_f Xf_{eq,f}\left(1-f_{eq,f}\right)}{T(u\cdot p)}},\\
		{c_3} =& {0}.
	\end{align}
	Similarly, the coefficients $\bar{c}_1, \bar{c}_2$ and 
	$\bar{c}_3$  
	associated to antiquarks can be evaluated.
	The general form of bulk viscous pressure is given as
	\begin{align}\label{bulk_viscous}
		\Pi = \zeta^{\mu\nu}\partial_{\mu}u_{\nu},
	\end{align} 
	where the tensor coefficient $\zeta^{\mu\nu}$ is known as bulk 
	viscosity coefficient. $\zeta^{\mu\nu}$ can be expressed in terms 
	of projection basis as
	\begin{equation}\label{general}
		\zeta^{\mu\nu} = \zeta_1 A_{(1)}^{\mu\nu} + \zeta_2 A_{(2)}^{\mu\nu} + \zeta_3 A_{(3)}^{\mu\nu},
	\end{equation}
	where $\zeta_1, \zeta_2$ and $\zeta_3$ is the longitudinal, transverse 
	and Hall component of bulk viscosity respectively. Using the 
	integral form of bulk viscous pressure 
	\begin{equation}
		{\Pi =\frac{\Delta^{\mu\nu}}{3}\int dP~p^{\mu}p^{\nu}\left(\delta f+\delta\bar{f}\right),}
	\end{equation}
	we obtained $\zeta_1, \zeta_2$ and $\zeta_3$ as
	\begin{align}\label{bulk}
		{\zeta_1 = \zeta_2 = \frac{\Delta^{\mu\nu}}{3}\int dP~ p_{\mu}p_{\nu}~c_1},
	\end{align}
	and
	\begin{equation}
		\zeta_3 = 0.
	\end{equation}
	Since, the dissipative part of energy-momentum tensor, $\Delta 
	T^{\mu\nu}$ and the particle current, $n^{\mu}$ cannot be determined 
	uniquely by the second law of
	thermodynamics, one usually introduces some constraints to
	fix them, they are known as matching conditions. The matching (or 
	fitting) conditions are also introduced
	because of necessity of matching the energy density and
	baryonic charge density in non-equilibrium state to the corresponding 
	equilibrium densities, or equivalently,
	\begin{equation}
		u_{\mu}u_{\nu}\Delta T^{\mu\nu} =0;\qquad u_{\mu}\Delta N^{\mu} = 0.
	\end{equation}
	With the help of above matching conditions and using the definition of 
	$X$, Eq.\eqref{bulk} can be written as
	\begin{align}
		{\zeta_1=\zeta_2=\sum_{f}\frac{1}{3T}\int dP \frac{X^2}{\varepsilon_f}\tau_f\left[f_{eq,f}\left(1-f_{eq,f}\right)+\bar{f}_{eq,f}\left(1-\bar{f}_{eq,f}\right)\right]}.\label{b}
	\end{align}
	The longitudinal ($\zeta_1$) and transverse ($\zeta_2$) component of 
	bulk viscosity are same and the effect of magnetic field comes through 
	the thermal mass (squared) of quarks and antiquarks with magnetic 
	field correction. The above form of bulk viscosity is qualitatively similar to the Ref. \cite{ashutosh,shubha}, where difference arises due to the mass. We have so far determined how quarks and antiquarks 
	contribute to bulk viscosity. We shall now determine the gluonic 
	contribution to bulk viscosity. Since, gluons do not interact with 
	electromagnetic field, the Boltzmann transport equation has the form 
	as follows
	\begin{equation}\label{gluon_boltz}
		p^{\mu}\partial_{\mu}f_{eq,g} = \frac{-u\cdot p}{\tau}\delta f_g.
	\end{equation}
	Using the Eq.\eqref{dT} and taking gluon chemical potential to be 
	zero, the left-hand side of Eq.\eqref{gluon_boltz} can be expressed as 
	follows,
	\begin{align}
		p^{\mu}\partial_{\mu}f_{eq,g}=&\left(\frac{\partial f_{eq,g}}{\partial T}\right)\Big[\left(u\cdot p\right)DT + p^{\mu}\nabla_{\mu}T\Big] + 	\left(\frac{\partial f_{eq,g}}{\partial u^{\mu}}\right)\Big[\left(u\cdot p\right)Du^{\mu} + p^{\mu}\nabla_{\mu}u^{\nu}\Big],\\
		=&\frac{1}{T}f_{eq,g}\left(1+f_{eq,g}\right)\left[X'\nabla_{\mu}u^{\mu}-p^{\nu}p^{\mu}\left(\nabla^{\mu}u^{\nu}-\frac{1}{3}\Delta_{\mu\nu}\nabla_{\sigma}u^{\sigma}\right)\right],\label{97}
	\end{align}
	where $X' = 
	\left(u\cdot p\right)^2\left(\frac{4}{3}-\lambda'\right)-\frac{1}{3}m^2$. 
	On employing the Eq.\eqref{assum} to the right-hand side of 
	Eq.\eqref{gluon_boltz}, it can be written in terms of the basis of 
	projection tensor as
	\begin{equation}\label{98}
		{\frac{-u\cdot p}{\tau}\delta f_g =  \frac{-u\cdot p}{\tau}\Big[\left\{c_1\left(b_{\mu}b_{\nu}\right)+c_2\left(\Delta_{\mu\nu}-b_{\mu}b_{\nu}\right)+ic_3b_{\mu\nu}\right\}\left(\partial^{\mu}u^{\nu}\right)\Big]}.
	\end{equation}
	On equating the coefficients of $\Delta_{\mu 
		\nu}\partial^{\mu}u^{\mu}, 
	b^{\mu}b^{\nu}\partial_{\mu}u_{\nu}$ and 
	$b^{\mu\nu}\partial_{\mu}u_{\nu}$ on both sides of Eq.\eqref{97} and 
	\eqref{98}, we obtain
	\begin{equation}
		{c_2 = -\frac{\tau}{(u\cdot p)T}\left[(u.p)^2\left(1-\lambda'\right)\right]f_{eq,g}\left(1+f_{eq,g}\right)}.
	\end{equation}
	Hence, the bulk viscosity coefficient for gluons is obtained to be as
	\begin{equation}
		{\zeta_g = \frac{\tau_g}{T}\int dP\frac{X^{'2}}{\varepsilon_g^2}f_{eq,g}\left(1+f_{eq,g}\right)}.
	\end{equation}
	This gluonic contribution of bulk viscosity ($\zeta_g$) will be added 
	to the fermionic contribution of bulk viscosity ($\zeta_1, \zeta_2$).
	\section{Results and Discussions}\label{results}
	We have summarized the results for the anisotropic components of shear 
	and bulk viscosity coefficients in RTA for non-zero magnetic field and 
	quark chemical potential.
	\subsection{Shear Viscosity}
	\begin{figure}
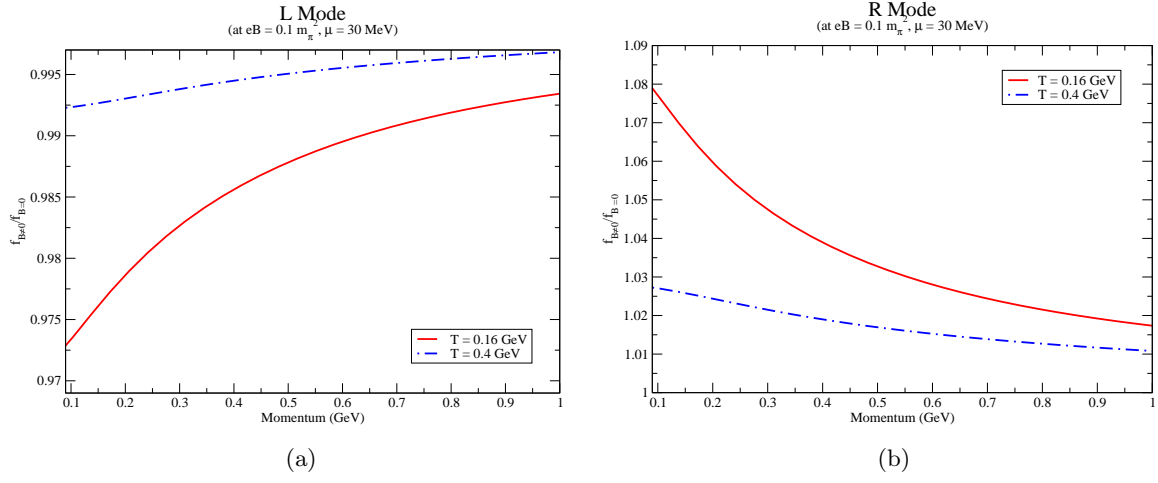

		\begin{subfigure}{0.48\textwidth}
			\includegraphics[width=0.95\textwidth]{dist_func_L_ratio_p.eps}
			\caption{}
		\end{subfigure}
		\begin{subfigure}{0.48\textwidth}
			\includegraphics[width=0.95\textwidth]{dist_func_R_ratio_p.eps}
			\caption{}
		\end{subfigure}
		\caption{Variation of $f_{B\neq 0}/f_{B=0}$ for L mode (a) and R mode (b) with momentum at low and high temperature.}\label{dist_p}
	\end{figure}
	The interaction among partons has been incorporated via the quasiparticle mass. Thus, in order to study the results on momentum transport, we need to understand the effects of $m_{L/R}^2$ on the quark distribution function which is shown in Fig. \eqref{dist_p}. $f_{B\neq 0}$ for L and R mode is defined using the mass $m_{L}^2$ and $m_{R}^2$ whereas $f_{B=0}$ is defined using $m_{th}^2$.
	As can be seen from the figures, this ratio is found to be smaller than unity for the L mode and larger than unity for the R mode, which suggests that the presence of a magnetic field causes a decrease in the L mode distribution function, and an increase in the R mode distribution function, compared to the distribution function at $B=0$. This effect of mass will also be reflected in the magntiude of shear and bulk viscosity.
	
	\begin{figure}
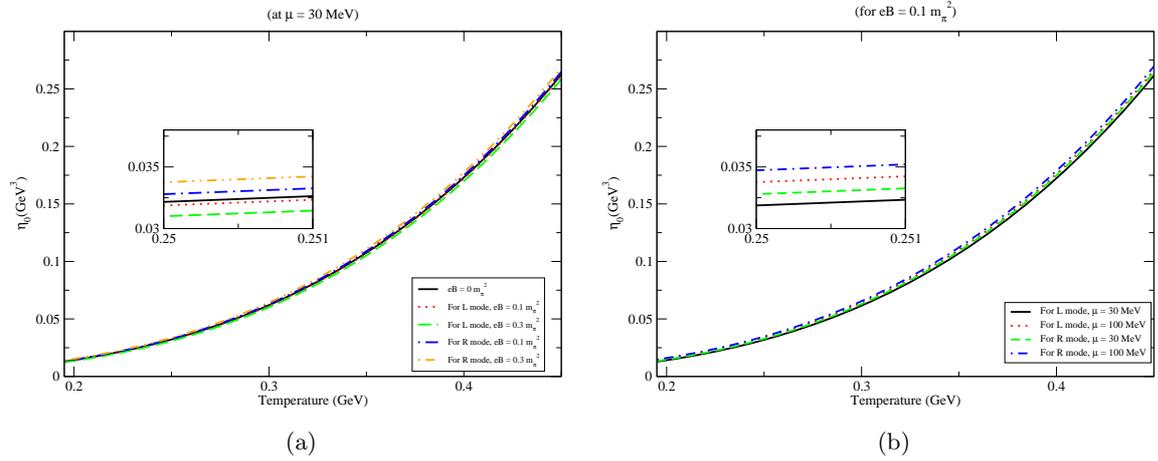

		\begin{subfigure}{0.48\textwidth}
			\includegraphics[width=0.95\textwidth]{shear_0_l_r.eps}
			\caption{}
		\end{subfigure}
		\begin{subfigure}{0.48\textwidth}
			\includegraphics[width=0.95\textwidth]{shear_0_l_r_mu.eps}
			\caption{}
		\end{subfigure}
		\caption{Variation of $\eta_0$ with temperature at different fixed values of magnetic field (a) and quark chemical potential (b) for both L and R modes.}\label{shear_0}
	\end{figure}
	In Fig. \eqref{shear_0}, we have shown the variation of longitudinal 
	shear viscosity ($\eta_0$) with temperature at different fixed values of magnetic field and quark chemical potential for both L and R modes. For L mode, $\eta_0$ decreases with magnetic 
	field, whereas it increases for the R mode. This opposite behaviour is 
	a result of the different mass of the L/R mode, which increases/decreases with 
	magnetic field. The appearance of this mass in the denominator of 
	$\eta_0$, results in the decreasing/increasing behaviour of $\eta_0$ 
	with magnetic field. The L and R mode of $\eta_0$ shows an increment with $\mu$ due to the 
	increased number of quark number density. Further, the increasing 
	behaviour of $\eta_0$ for both modes with temperature is attributed to 
	the Boltzmann factor $\exp(-\varepsilon(p)/T)$ in the distribution 
	function. $\eta_0$ will have nonvanishing contribution for zero 
	magnetic field and quark chemical potential.
	\begin{figure}
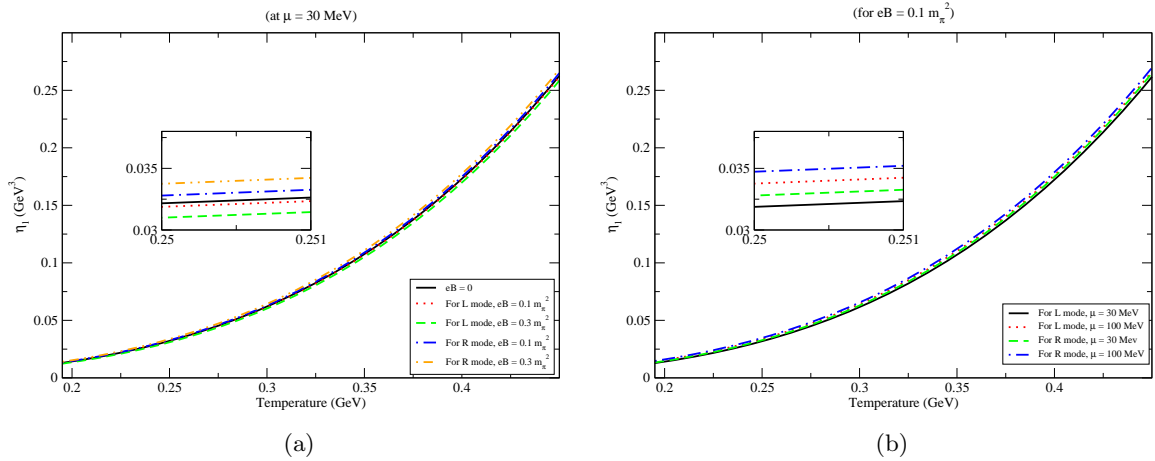

		\begin{subfigure}{0.48\textwidth}
			\includegraphics[width=0.95\textwidth]{shear_1_l_r.eps}
			\caption{}
		\end{subfigure}
		\begin{subfigure}{0.48\textwidth}
			\includegraphics[width=0.95\textwidth]{shear_1_l_r_mu.eps}
			\caption{}
		\end{subfigure}
		\caption{Variation of $\eta_1$ with temperature at different fixed values of magnetic field (a) and quark chemical potential (b) for both L and R modes.}\label{shear_1_mu}
	\end{figure}
	Fig. \eqref{shear_1_mu} shows the variation of 
	$\eta_1$ (transverse component of shear viscosity)  with temperature 
	at 
	different fixed values of magnetic field and quark chemical potential 
	respectively. Similar to the $\eta_0$, $\eta_1$ decreases for L mode 
	and increases for R mode with magnetic field and increases with quark 
	chemical potential for both modes. In the absence of magnetic field, 
	transverse component becomes equal to the longitudinal component. Also it possesses the finite 
	contribution at $\mu = 0$ due to the equal contribution from quarks 
	and 
	antiquarks in same direction.
	
	\begin{comment}
		\begin{figure}[H]
			\begin{subfigure}{0.48\textwidth}
				\includegraphics[width=0.95\textwidth]{dist_func_L_ratio_T.eps}
				\caption{}
			\end{subfigure}
			\begin{subfigure}{0.48\textwidth}
				\includegraphics[width=0.95\textwidth]{dist_func_R_ratio_T.eps}
				\caption{}
			\end{subfigure}
			\caption{Variation of $f_{B\neq 0}/f_{B=0}$ for L mode (a) and R mode (b) with temperature at low and high momentum.}\label{dist}
		\end{figure}
	\end{comment}
	
	\begin{figure}
		\centering
		\includegraphics[width=0.48\textwidth]{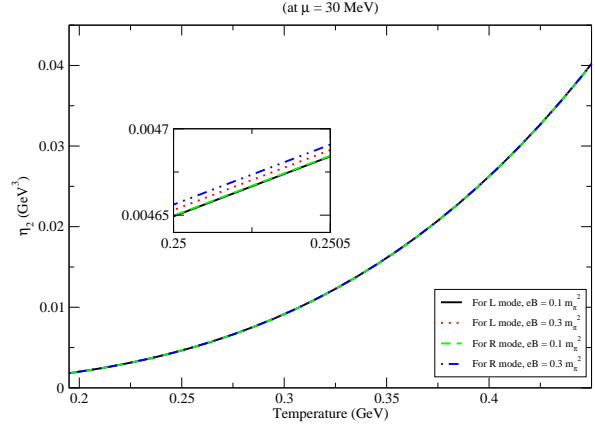}
		\caption{Variation of $\eta_2$ with temperature at different fixed values of magnetic field for both L and R modes.}\label{shear_2}
	\end{figure}

	\begin{figure}
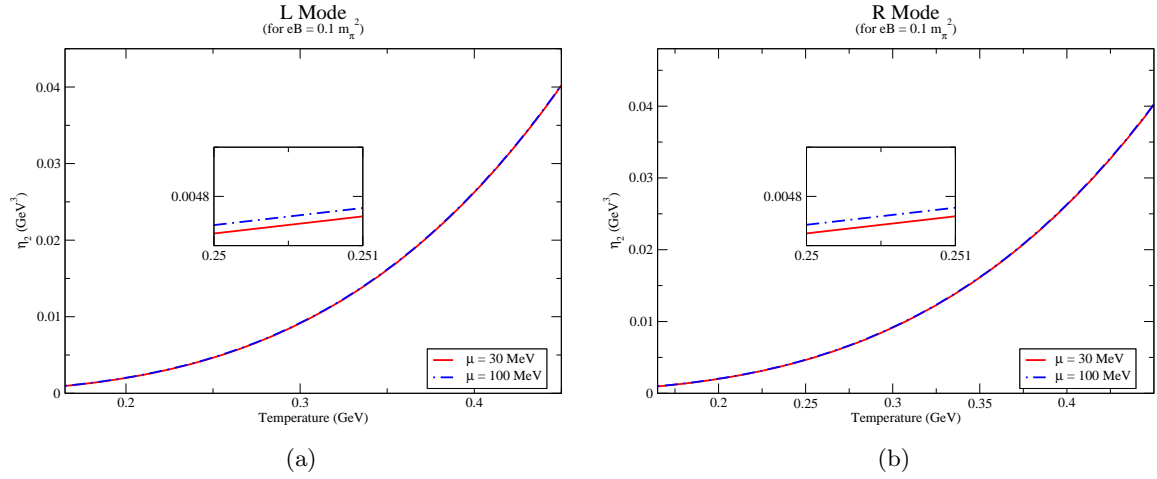

		\begin{subfigure}{0.48\textwidth}
			\includegraphics[width=0.95\textwidth]{shear_2_l_mu.eps}
			\caption{}
		\end{subfigure}
		\begin{subfigure}{0.48\textwidth}
			\includegraphics[width=0.95\textwidth]{shear_2_r_mu.eps}
			\caption{}
		\end{subfigure}
		\caption{Variation of $\eta_2$ for L mode (a) and R mode (b) with 
			temperature at different fixed values of quark chemical 
			potential.}\label{shear_2_mu}
	\end{figure}
	Figs. \eqref{shear_2} and \eqref{shear_2_mu} shows the variation of 
	$\eta_2$ (Hall component of shear viscosity) with temperature at 
	different fixed values of magnetic field and quark chemical potential 
	respectively. Since, $\eta_2$ is proportional to $qB$, therefore it 
	increases with magnetic field for both modes. Further, its magnitude 
	amplifies with quark chemical potential. This Hall-type shear 
	viscosity 
	vanishes in the absence of magnetic field. At zero chemical potential, 
	the number of particles and antiparticles are same and they will give 
	equal and opposite contribution to the shear viscosity component, 
	$\eta_2$. Hence, it vanishes for zero quark chemical potential even in 
	the ambience of nonvanishing magnetic field. 
	\begin{figure}[H]
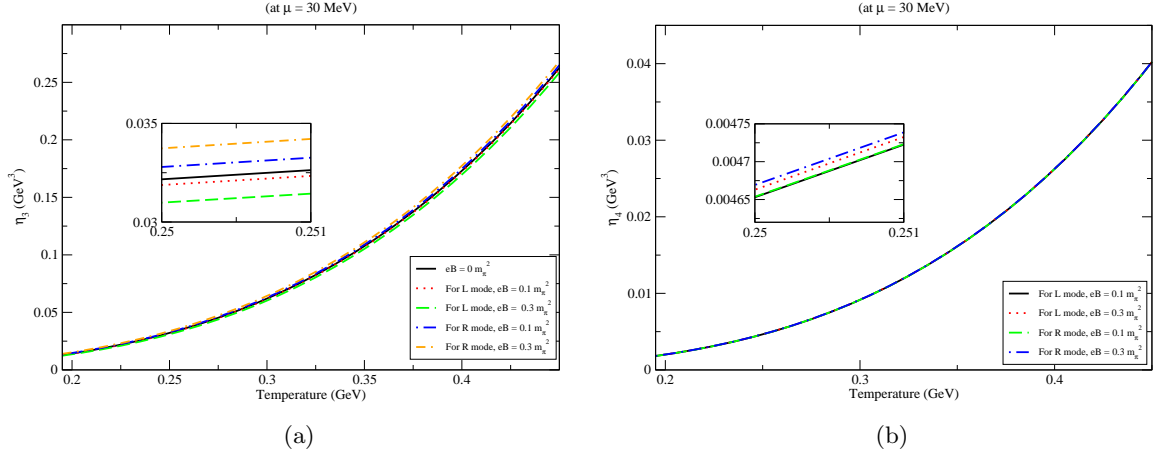

		\begin{subfigure}{0.48\textwidth}
			\includegraphics[width=0.95\textwidth]{shear_3_l_r.eps}
			\caption{}
		\end{subfigure}
		\begin{subfigure}{0.48\textwidth}
			\includegraphics[width=0.95\textwidth]{shear_4_l_r.eps}
			\caption{}
		\end{subfigure}
		\caption{Variation of $\eta_3$ (a) and $\eta_4$ (b) with temperature for both L and R modes at different fixed values of magnetic 
			field.}\label{shear_3}
	\end{figure} 
	The variation of another transverse component of shear viscosity 
	($\eta_3$) and Hall component ($\eta_4$) is shown in Fig. \eqref{shear_3}. 
	Similar to the $\eta_1$, 
	$\eta_3$ decreases/increases with magnetic field for L/R mode of 
	quarks. In the absence of magnetic field, $\eta_1$ and $\eta_3$ 
	becomes equal to the longitudinal component, i.e. $\eta_0 = \eta_1 = 
	\eta_3$. The same as $\eta_0$ and $\eta_1$, $\eta_3$ will have a finite 
	contribution at the vanishing quark chemical potential.
	The
	another Hall component of shear viscosity ($\eta_4$) shows the same behaviour as $\eta_2$ with temperature and magnetic field. It also 
	vanishes for zero magnetic field and for symmetric quark chemical 
	potential at finite magnetic field. The increasing behaviour of all the coefficients with temperature is due to the Boltzmann factor $\exp(-\varepsilon(p)/T)$. They also exhibit an increase with quark chemical potential, similar to the other shear viscous coefficients. 
	\subsection{Bulk Viscosity}
	\vspace{5mm}
	\begin{figure}[H]
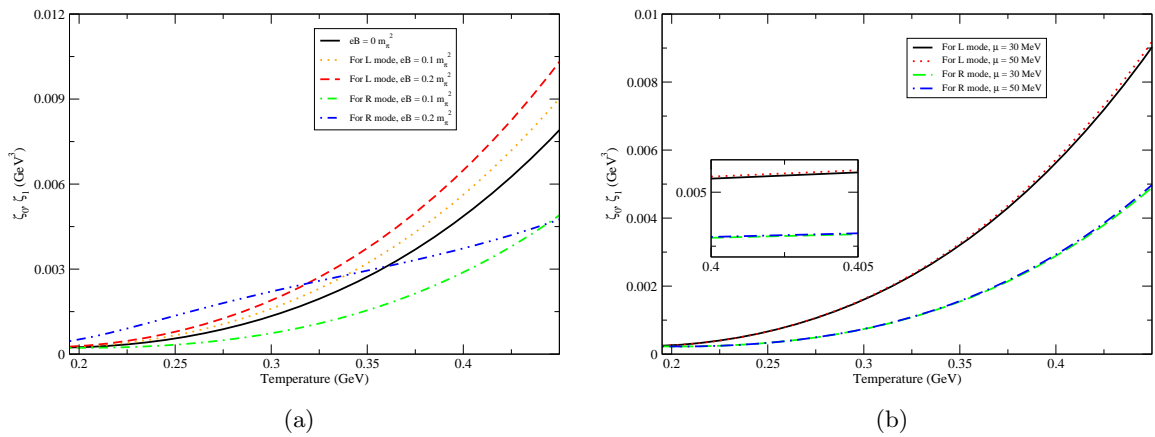

		\begin{subfigure}{0.48\textwidth}
			\includegraphics[width=0.95\textwidth]{bulk_l_r.eps}
			\caption{}
		\end{subfigure}
		\begin{subfigure}{0.48\textwidth}
			\includegraphics[width=0.95\textwidth]{bulk_l_r_mu.eps}
			\caption{}
		\end{subfigure}
		\caption{Variation of $\zeta_0, \zeta_1$ for L mode (a) and R mode (b) 
			with temperature at different fixed values of magnetic 
			field.}\label{bulk_vis}
	\end{figure} 
	The variation of longitudinal and transverse bulk viscosity ($\zeta_0$ 
	and $\zeta_1$) with temperature at different fixed values of magnetic 
	field and quark chemical potential is shown in Fig. \eqref{bulk_vis}. Both L and R mode bulk viscosity have higher magnitude at $eB = 0.2 m_{\pi}^2$ than at $eB = 0.1 m_{\pi}^2$. This behaviour is attributed due to the factor, $X=(u\cdot p)^2\left(\frac{4}{3}-\lambda'\right)-\frac{1}{3}m^2+\left(u\cdot p\right)\left[\left(\lambda''-1\right)h-\lambda'''T\right]$. Further, the Boltzmann factor $f_{eq}(1-f_{eq})$ leads to 
	the increasing behaviour of bulk viscosity with temperature. The 
	increment of bulk viscosity with quark chemical potential for both 
	modes is due to the increased contribution from quarks than antiquarks. 
	$\zeta_0$ and $\zeta_1$ will have finite contribution at zero quark 
	chemical potential even in the absence of magnetic field.
	\section{Applications}\label{applications}
	In this part, we will examine the influence of a weak magnetic field on 
	the specific shear and bulk viscosity, and Reynolds number.
	\subsection{Specific shear and bulk viscosities}
	The specific shear ($\eta/s$) and bulk viscosity ($\zeta/s$) are the 
	viscosity to entropy density ratio. A small value for $\eta/s$ was 
	estimated from the analysis of the elliptic flow data \cite{paul} and 
	was found to be close to the conjectured lower bound $\eta/s|_{KSS}$ 
	from AdS/CFT \cite{PhysRevLett.87.081601}. This led to claim that the 
	QGP formed at RHIC was the most
	perfect fluid ever observed.
	%The 
	%small value of $\eta/s$ has also been reported for QGP using parton 
	%cascade approach suggesting that the matter produced at the RHIC is 
	%strongly coupled fluid of quarks and gluons 
	%\cite{PhysRevLett.101.082302,FERINI2009325,CASSING2009215}. The 
	%relativistic viscous hydrodynamics model \cite{paul} was also used to 
	%determine the $\eta/s$ and was compared with experimental \cite{aziz} 
	%and lattice result \cite{nakamura}. 
	We have plotted the five  specific 
	shear viscous and two specific bulk viscous coefficients.\par
	\begin{figure}
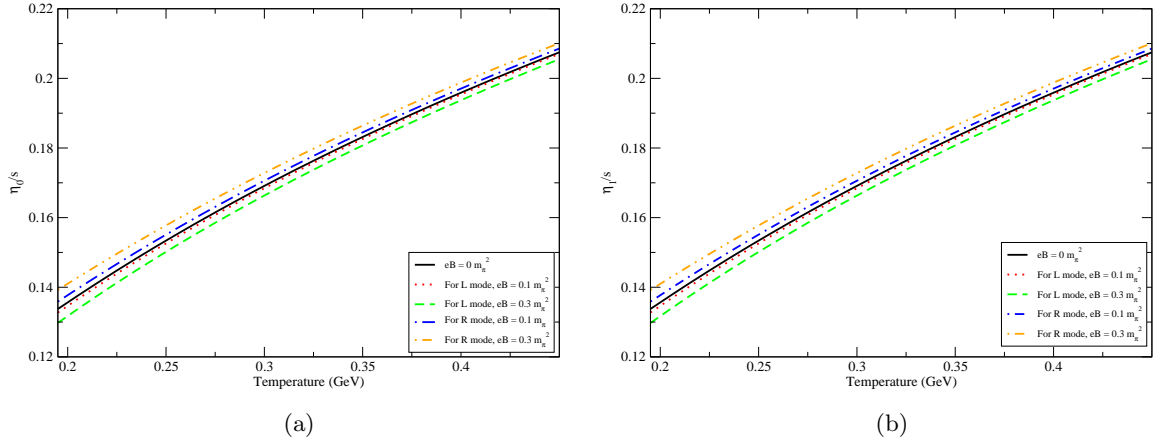

		\begin{subfigure}{0.48\textwidth}
			\includegraphics[width=0.95\textwidth]{shear_0_ratio_l_r.eps}
			\caption{}
		\end{subfigure}
		\begin{subfigure}{0.48\textwidth}
			\includegraphics[width=0.95\textwidth]{shear_1_ratio_l_r.eps}
			\caption{}
		\end{subfigure}
		\caption{Variation of $\eta_0/s$ (a) and $\eta_1/s$ (b) with temperature for both L and R mode at 
			different fixed values of magnetic field.}\label{ss_0}
	\end{figure}
	\begin{figure}[H]
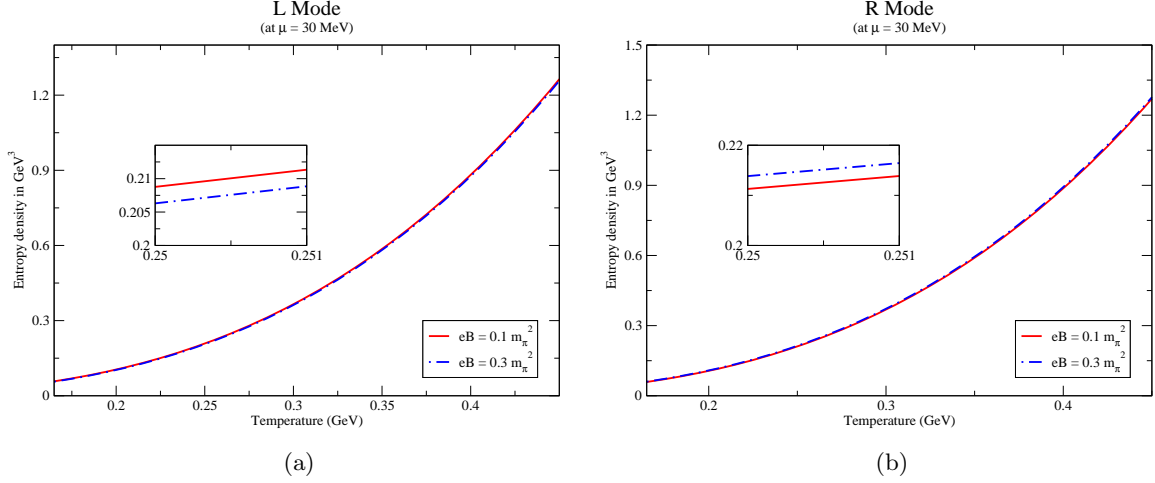

		\begin{subfigure}{0.48\textwidth}
			\includegraphics[width=0.95\textwidth]{entropy_l.eps}
			\caption{}
		\end{subfigure}
		\begin{subfigure}{0.48\textwidth}
			\includegraphics[width=0.95\textwidth]{entropy_r.eps}
			\caption{}
		\end{subfigure}
		\caption{Variation of entropy density for L mode (a) and R mode (b) at 
			different fixed values of magnetic field.}\label{entrop}
	\end{figure}
	The variation of $\eta_0/s$ and $\eta_1/s$ with temperature 
	at different fixed values of magnetic field is shown in 
	Fig. \eqref{ss_0}.  The entropy density for L mode decreases and increases 
	for R mode with magnetic field as shown in Fig. \eqref{entrop}. 
	This 
	behaviour of entropy density with 
	$B$ is attributable to the Boltzmann factor $f_{eq}(1-f_{eq})$.
	The rate of change of magnitude for shear viscosity with magnetic field 
	is higher than the entropy density, hence $\eta_0/s$ and $\eta_1/s$ 
	shows the same behaviour as $\eta_0$ and $\eta_1$ with magnetic field.
	
	\vspace{5mm}

	\begin{figure}[H]
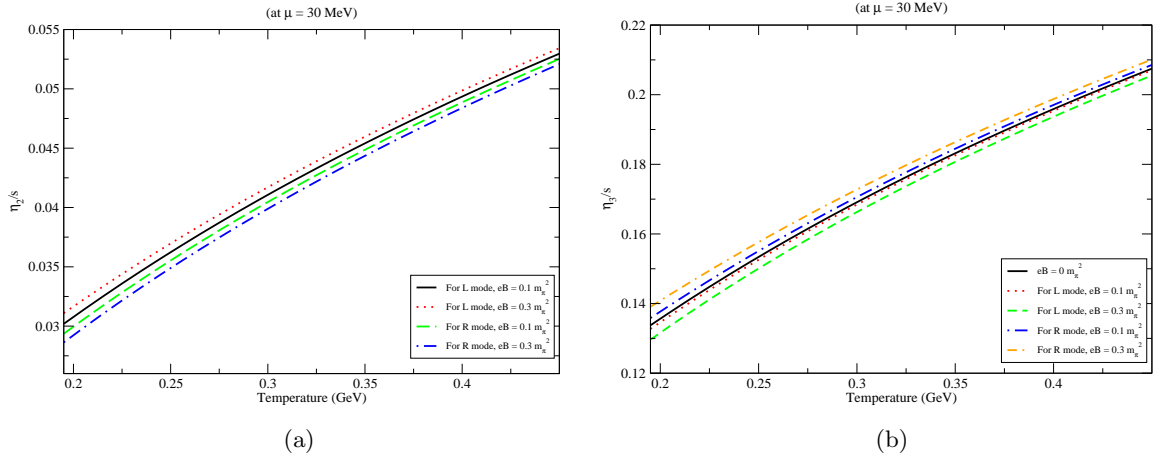

		\begin{subfigure}{0.48\textwidth}
			\includegraphics[width=0.95\textwidth]{shear_2_ratio_l_r.eps}
			\caption{}
		\end{subfigure}
		\begin{subfigure}{0.48\textwidth}
			\includegraphics[width=0.95\textwidth]{shear_3_ratio_l_r.eps}
			\caption{}
		\end{subfigure}
		\caption{Variation of $\eta_2/s$ (a) and $\eta_3/s$ (b) with temperature at 
			different fixed values of magnetic field for both L and R modes.}\label{ss_1}
	\end{figure}
	\begin{figure}[H]
		\centering
		\includegraphics[width=0.48\textwidth]{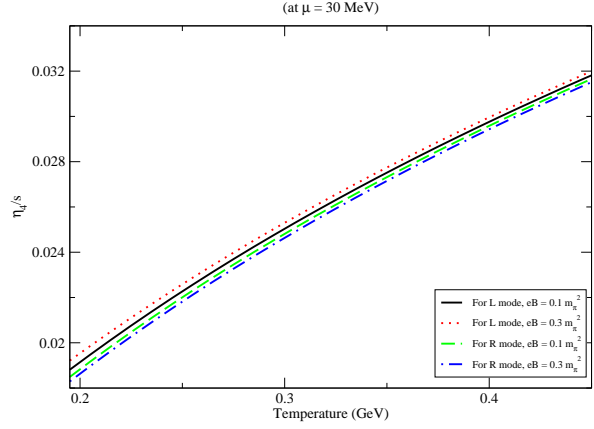}
		\caption{Variation of $\eta_4/s$ with temperature for L 
			and R mode at different fixed values of magnetic field.}\label{ss_4}
	\end{figure}
	The behaviour of $\eta_2/s$ and $\eta_3/s$ with temperature is shown in Fig. \eqref{ss_1}. The behaviour of $\eta_2/s$ for L mode is same as $\eta_2$ but 
	opposite for R mode. The rate of change of magnitude of entropy density with magnetic field 
	is higher than the $\eta_2$, hence entropy density determines the 
	trend 
	of $\eta_2/s$. Whereas the rate of change of magnitude of $\eta_3$ with magnetic field is higher than the entropy density, hence $\eta_3/s$ shows the same behaviour as $\eta_3$.
	The variation of $\eta_4/s$ 
	with 
	temperature is shown in Fig. \eqref{ss_4} behaving in 
	the same manner $\eta_2/s$. We observed that the shear viscosity to entropy density ratio for longitudinal and transverse 
	components in weak field limit is greater than $\frac{1}{4\pi}$ as 
	expected because in absence of magnetic field, this ratio is higher 
	than 
	KSS lower bound. The Hall component of shear viscosity vanishes in the 
	absence of a magnetic field and its magnitude is very small as 
	compared 
	to the longitudinal and transverse components. The corresponding ratio 
	for Hall component is
	found to be less than $\frac{1}{4\pi}$.
	
	\vspace{5mm}
	\begin{figure}[H]
		\centering
		\includegraphics[width=0.48\textwidth]{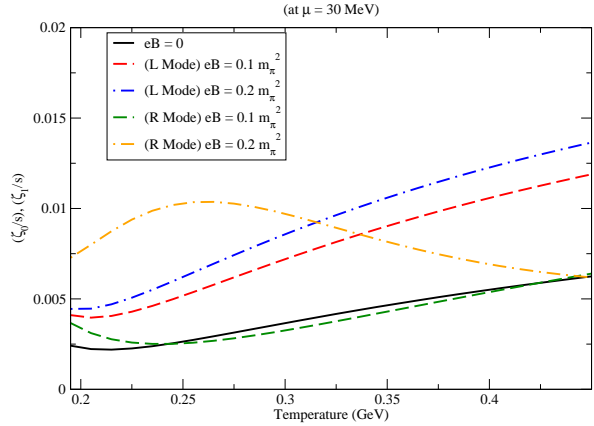}
		\caption{Variation of $\zeta_0/s$, $\zeta_1/s$ with temperature for L 
			and R mode at different fixed values of magnetic field.}\label{bulk_ratio}
	\end{figure}
	The variation of bulk viscosity to entropy density ratio, $\zeta_0/s$ 
	and $\zeta_1/s$, with 
	temperature at different fixed values of magnetic field is shown in 
	Fig. \eqref{bulk_ratio}. Similar to the $\zeta_0$ and $\zeta_1$, the specific 
	bulk viscosities have greater magnitude at $eB = 0.2 m_{\pi}^2$ 
	compared to $eB = 0.1 m_{\pi}^2$.
	The $\zeta_0/s$ and $\zeta_1/s$ were found to be less than shear 
	viscosity to entropy density ratio. Both $\zeta_0/s$ and $\zeta_1/s$ 
	for L and R mode exhibits the non-monotonic behaviour with 
	temperature. 
	The ratio for L mode shows a minimum at $T$=0.205 GeV for $eB = 0.1 
	m_{\pi}^2$ and at $T$ = 0.195 GeV for $eB = 0.2 m_{\pi}^2$. The R mode 
	ratio for $eB= 0.1 m_{\pi}^2$ has a minimum at $T$=0.235 GeV and 
	maximum 
	at $T$=0.265 GeV for $eB = 0.2 m_{\pi}^2$. This non-monotonic 
	behaviour 
	has also been observed in the estimations based on NJL model 
	\cite{paramita}, linear sigma model at large N limit \cite{antonio}. 
	The 
	non-zero value of bulk viscosity suggests the departure from 
	conformality. 
	\subsection{Reynolds number}
	Reynolds number ($\text{Rl}$) is one of the most important 
	dimensionless quantities in microfluidics.
	$\text{Rl}$ is significant for characterizing a fluid's transport 
	characteristics and for figuring out the type of flow pattern a fluid 
	exhibits. It is defined in terms of characteristic length ($L$), 
	relative velocity of the fluid ($v$) and kinematic viscosity 
	($\eta/\rho$) as
	\begin{equation}
		\text{Rl} = \frac{Lv}{\eta/\rho}.
	\end{equation}
	The small value of $\text{Rl} $ suggests the laminar flow, which 
	characterizes the more viscous fluid, while its large value, typically 
	in thousands, indicates the turbulent flow.
	
	\vspace{5mm}
	\begin{figure}[H]
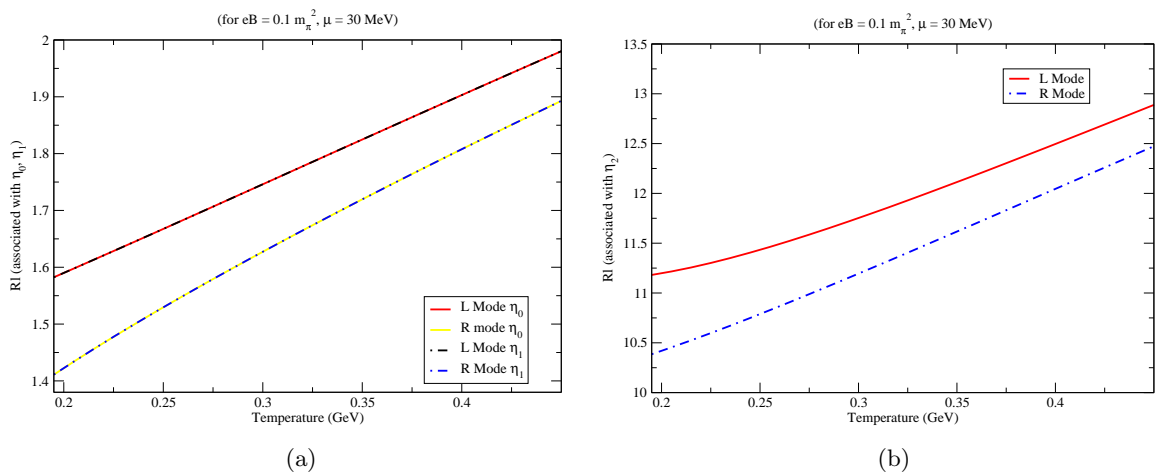

		\begin{subfigure}{0.48\textwidth}
			\includegraphics[width=0.95\textwidth]{reynold.eps}
			\caption{}
		\end{subfigure}
		\begin{subfigure}{0.48\textwidth}
			\includegraphics[width=0.95\textwidth]{reynold_hall.eps}
			\caption{}
		\end{subfigure}
		\caption{Variation of Reynolds number associated with longitudinal, 
			transverse (a) and Hall (b) component of shear viscosity, with 
			temperature.}\label{reynold}
	\end{figure}
	Fig. \eqref{reynold} shows 
	the variation of Reynolds number associated with $\eta_0, \eta_1$ and 
	$\eta_2$, with temperature at $eB = 0.1 m_{\pi}^2$ and $\mu = 30$ MeV.  
	The mass density for L mode is higher than R mode whereas the magnitude 
	of shear viscosity for L mode is lower than R mode. Therefore, Rl 
	associated with $\eta_0, \eta_1$ and $\eta_2$ for L mode has higher 
	magnitude than for R mode. Further, Rl associated with $\eta_0$ and 
	$\eta_1$ has the same magnitude in their respective modes. 
	Csernai \textit{et al}. estimated the value of $\text{Rl}$ in the range 
	between 3 to 10 for initial QGP with $\eta/s = 0.1$ using a 
	(3+1)-dimensional fluid dynamical model \cite{cse}. Moreover, an upper 
	bound on Reynolds number has been investigated from holographic point of 
	view for nearly central collisions at given temperature and also at 
	different values of magnetic field \cite{brett}.

	\section{Conclusions}\label{conc}
	In this work, we aimed to investigate the impact of weak magnetic 
	field and 
	baryon asymmetry on shear and bulk viscosities in a hot QCD medium, using kinetic theory 
	under RTA. In a weak
	magnetic field, we have found that the L and R chiral modes of quarks 
	get
	separated due to difference in their mass and become non-degenerate 
	contrary to the
	strong magnetic field case. Moreover, the introduction of magnetic 
	field breaks the isotropy of the 
	medium, resulting in a decomposition of shear and bulk viscosity into 
	distinct components. The shear viscosity is decomposed into five 
	components, including the longitudinal 
	component ($\eta_0$), transverse components ($\eta_1$ and $\eta_3$) 
	and 
	Hall components ($\eta_2$ and $\eta_4$). In the 
	absence of magnetic field, the transverse components assume the same 
	form 
	as longitudinal component, while the Hall components vanish for both 
	modes. The 
	decrease or increase of $\eta_0, \eta_1$ and $\eta_3$ with the 
	magnetic 
	field for L or R mode is attributed to the different values of 
	effective 
	quark mass for both modes. On the other hand, the increase of $\eta_2$ 
	and $\eta_4$ with magnetic field is due to the direct dependence of 
	magnetic field on Hall components. The longitudinal, transverse and 
	Hall components positively amplify with the quark chemical potential 
	for 
	both L and R modes. In the absence of quark chemical potential, the 
	Hall component for the shear viscosity vanishes for both modes, even 
	in the presence of a magnetic field. The bulk viscosity is decomposed 
	into two components, including the
	longitudinal ($\zeta_0$) and transverse ($\zeta_1$) component. Both 
	$\zeta_0$ and $\zeta_1$ have higher magnitude at $eB = 0.2 m_{\pi}^2$ 
	than at $eB= 0.1 m_{\pi}^2$ in both L and R mode. Similar to the shear 
	viscosity, bulk viscosity also 
	increases with quark chemical potential. The shear viscosity to 
	entropy density ratio for longitudinal ($\eta_0/s$) and transverse 
	components ($\eta_1/s$, $\eta_3/s$) exhibit the decrease for L mode 
	and increase for R mode as the magnetic field intensity increases. 
	Furthermore, the Hall component of specific shear viscosity increases 
	with magnetic field for L mode and decreases for R mode. Moreover, the 
	ratios of shear viscosity to entropy density are found to be greater 
	than $1/(4\pi)$ for longitudinal and transverse components,
	whereas the ratio is less than $1/(4\pi)$ for Hall component. The bulk 
	viscosity to entropy density ratios ($\zeta_0/s$ and $\zeta_1/s$) show 
	an increase with magnetic field for both L and R modes, and these 
	ratios
	exhibit non-monotonic behaviour with temperature, showing minima or 
	maxima around $T\sim$ 200 MeV.
	%Pr has found to be greater than unity for both 
	%modes thus suggesting the dominance of momentum diffusion over the 
	%thermal diffusion in sound attenuation. 
	Additionally, the Rl for L mode is higher than the R mode due to the 
	different mass densities.
	The small value of Rl suggests 
	the more viscous fluid, hence describing the laminar flow.
	
	\section*{Acknowledgements}
	P. P. would like to thank Najmul Haque for helpful discussions, and Debarshi Dey and Salman Ahamad Khan for a critical reading of the manuscript.
	
	\appendix
	\section{ Calculation of structure functions}\label{append}
	Here, we will show the computation of structure functions from Eq.\eqref{A} to \eqref{D} in one-loop order for hot and weakly magnetized medium under HTL approximation.
	Since, trace of odd number of gamma matrices is zero, the Eq.\eqref{A} can be written as
	\begin{equation}
		\mathcal{A} = \frac{1}{4}\frac{\text{Tr}(\Sigma_{0}\slashed{P})- (P.u)\text{Tr}(\Sigma_{0}\slashed{u})}{(P\cdot u)^2-P^2},
	\end{equation}
	where, 
	\begin{equation}
		\Sigma_{0} = g^2 C_F T\sum_{n}\int\frac{d^3 k}{(2\pi)^3}\gamma_{\mu}\frac{\slashed{K}}{K^2-m_f^2}\gamma^{\mu}\frac{1}{(P-K)^2}.
	\end{equation}
	Using the following two traces:
	\begin{align}
		&\text{Tr}\left[\gamma_{\mu}\slashed{K}\gamma^{\mu}\slashed{P}\right] = -8K\cdot P,\\
		&\text{Tr}\left[\gamma_{\mu}\slashed{K}\gamma^{\mu}\slashed{u}\right] = -8K\cdot u,\\
	\end{align}
	we obtain,
	\begin{equation}
		\mathcal{A}(P) = \frac{1}{4\bf{|p|}^2}g^2 C_F\left[I_1(P)+I_2(P)\right],
	\end{equation}
	where $(P\cdot u)^2-P^2 = \bf{|p|}^2$. We will use the frequency sum to evaluate $I_1(P)$ and $I_2(P)$ with $k_0 = i\omega_n$, $p_0 = i\omega$, $E_1 = \sqrt{k^2+m_{f0}^2}$ and $E_2 = \sqrt{(p-k)^2}$. The frequency sum for fermion-boson case is \cite{bellac}
	\begin{equation}
		T\sum_{n}\tilde{\Delta}_{s1}(i\omega_n,E_1)\Delta_{s2}(i(\omega-\omega_n),E_2) =\sum_{s1,s2=\pm 1} -\frac{s_1 s_2}{4E_1E_2}\frac{\left(1-\tilde{f}(s_1E_1)+f(s_2E_2)\right)}{i\omega-s_1E_1-s_2E_2}.
	\end{equation}
	The leading $T^2$ behaviour will come from $s_1 = -s_2 = 1$ with $E_1\approx k$ and $E_2= |\bf{p-k}|$. Defining light-like four-vector $\hat{K}=(-i,\bf{\hat{k}})$ and $\hat{K}^{\prime} = (-i,\bf{-\hat{k}})$, we have,
	\begin{align}
		&i\omega + E_1 - E_2\simeq i\omega + {\bf{p\cdot\hat{k}}} = P\cdot\hat{K},\\
		&i\omega - E_1 + E_2\simeq i\omega - {\bf{p\cdot\hat{k}}} = P\cdot\hat{K^{\prime}},
	\end{align}
	and using the angular integration under HTL approximation,
	\begin{equation}
		\int\frac{d\Omega}{4\pi}\frac{\hat{K}\cdot u}{P\cdot\hat{K}} = \frac{1}{|\bf{p}|}Q_0\left(\frac{p_0}{|\bf{p}|}\right),
	\end{equation}
	we get,
	\begin{equation}
		\mathcal{A}(p_0,{|\bf{p}}|) = \frac{m_{th}^2}{{|\bf{p}}|^2}Q_1\left(\frac{p_0}{|\bf{p}|}\right).
	\end{equation} 
	Similarly, structure function $\mathcal{B}$ can be evaluated as
	\begin{equation}
		\mathcal{B}(p_0,{|\bf{p}}|) =- \frac{m_{th}^2}{{|\bf{p}|}^2}\left[\frac{p_0}{|\bf{p}|}Q_1\left(\frac{p_0}{|\bf{p}|}\right) - Q_0\left(\frac{p_0}{|\bf{p}|}\right)\right].
	\end{equation} 
	Using Eq.\eqref{self_energy} in \eqref{C} and \eqref{D}, where the contribution from $\Sigma_{0}$ vanishes due to the trace of odd no. of gamma matrices and we get the non-vanishing contribution form $\Sigma_{1}$ only and hence we get,
	\begin{align}
		&\mathcal{C}\left(p_0,{\bf{|p|}}\right)= -\frac{1}{4}\text{Tr}(\gamma_5\Sigma_{1}\slashed{u}),\\
		& \mathcal{D}\left(p_0, {\bf{|p|}}\right)= \frac{1}{4}\text{Tr}(\gamma_5\Sigma_{1}\slashed{b}).
	\end{align}
	Using the following two traces
	\begin{align}
		&\text{Tr}\left[\gamma_5\gamma_{\mu}\gamma_5\left[(K\cdot b)\slashed{u}-(K\cdot u)\slashed{b}\right]\gamma^{\mu}\slashed{u}\right] = 8(K\cdot b),\\
		&\text{Tr}\left[\gamma_5\gamma_{\mu}\gamma_5\left[(K\cdot b)\slashed{u}-(K\cdot u)\slashed{b}\right]\gamma^{\mu}\slashed{b}\right] = 8(K\cdot u),
	\end{align} 
	we obtain,
	\begin{align}
		&	\mathcal{C} = \frac{g^2 C_F|q_f B|}{4}T\sum_{n}\int\frac{d^3k}{(2\pi)^3}\frac{8 (K\cdot b)}{(K^2-m_{f0}^2)^2(P-K)^2},\\
		&	\mathcal{D} = -\frac{g^2 C_F|q_f B|}{4}T\sum_{n}\int\frac{d^3k}{(2\pi)^3}\frac{8 (K\cdot u)}{(K^2-m_{f0}^2)^2(P-K)^2},
	\end{align}
	which in turn requires the calculation of frequency sum \cite{ayala}
	\begin{align}
		&	Y = T\sum_{n}\Delta_F^2(K)\Delta_B(P-K), \\ \nonumber
		&	= \left(\frac{-\partial}{\partial m_{f0}^2}\right) T\sum_{n}\Delta_{F}(K)\Delta_{B}(P-K),
	\end{align}
	where,
	\begin{equation}
		T\sum_{n}\Delta_{F}(K)\Delta_{B}(P-K) = \sum_{s1,s2=\pm 1} -\frac{s_1 s_2}{4E_1E_2}\frac{\left(1-\tilde{f}(s_1E_1)+f(s_2E_2)\right)}{i\omega-s_1E_1-s_2E_2} .
	\end{equation}
	For $s_1 = -s_2 = 1$, we get,
	\begin{align}
		&	\mathcal{C} = \frac{4g^2C_F|q_fB|}{16\pi^2}\left(\frac{\pi T}{2m_{f0}} -\text{ln} 2 + \frac{7\mu^2\zeta(3)}{8\pi^2T^2}\right)\left[\frac{-p_z}{{\bf|p|}^2}Q_1\left(\frac{p_0}{|\bf{p}|}\right)\right],\\
		&	\mathcal{D} = -\frac{4g^2C_F|q_fB|}{16\pi^2}\left(\frac{\pi T}{2m_{f0}} -\text{ln} 2 + \frac{7\mu^2\zeta(3)}{8\pi^2T^2}\right)\left[\frac{1}{{\bf|p|}}Q_0\left(\frac{p_0}{|\bf{p}|}\right)\right].\\
	\end{align}
	\section{Derivation of Eq.\eqref{51}}\label{append_b}
	The equilibrium number density ($n = N^{\mu}u_{\mu}$) for system of 
	quarks for $f$th flavor is given by
	\begin{equation}\label{75}
		n = \sum_f g_f\int\frac{d^3p}{(2\pi)^3}\frac{p^\mu u_{\mu}}{p^0}f_{eq,f},
	\end{equation}	
	where $g_f= 3\times 2$ is the color and spin degeneracy factor. For  
	ease of the calculation, we introduce the two dimensionless quantities 
	as,	
	\begin{equation}
		y = \frac{m}{T},\qquad \xi = \frac{p^{\mu}u_{\mu}}{T} = \frac{1}{T}\left(\textbf{p}^2+m^2\right)^{1/2}.
	\end{equation}	
	In local rest frame, $\frac{d^3 p}{p^0}$ can be expressed as follows 
	using the abovementioned two quantities
	\begin{equation}
		\frac{d^3p}{p^0} = T^2\left(\xi^2-y^2\right)^{1/2}d\xi d\Omega,
	\end{equation}
	where $d\Omega = d(\text{cos}\theta)d\phi$ is the differential solid 
	angle. Using the expansion identity $\frac{1}{y^{-1}e^x-1} = 
	\sum_{k=1}^{\infty}\left(ye^{-x}\right)^k$, the integral Eq.\eqref{75} 
	can be expressed to a sum of the integral as 
	\begin{align}
		n =& \sum_f\frac{g_f}{2\pi^2}y^2T^3\sum_{k=1}^{\infty}e^{k\mu/T}k^{-1}K_2(ky)\\
		=& \sum_f\frac{g_f}{2\pi^2}y^2T^3 S^{1}_{2}(y),
	\end{align}
	where $S^{\alpha}_j(y$) = 
	$(\mp)^{k-1}\sum_{k=1}^{\infty}e^{k\mu/T}k^{-\alpha}K_j(ky)$ for 
	fermionic (bosonic), $K_{j}(y)$ is the modified Bessel function of 
	second kind of order $j$. Similarly, we can obtain the equilibrium 
	formulas for energy density ($en=u_{\mu}T^{\mu\nu}u_{\nu}$), pressure 
	($P= -\Delta_{\mu\nu}T^{\mu\nu}/3$) and enthalpy per particle ($h$) as
	\begin{align}
		en=&\sum_f g_f \int\frac{d^3p}{(2\pi)^3}\frac{\left(p^{\mu} u_{\mu}\right)^{2}}{p^0} f_{eq,f}=\sum_f\frac{g_fy^2T^4}{2\pi^2}\left[y S_{3}^{1}(y)-S_{2}^{2}(y)\right], \\\label{energy_density}
		P=&\sum_fg_f \int\frac{d^3p}{(2\pi)^3} \frac{p^{2}}{3p^0} f_{eq,f}=\sum_f\frac{g_f}{2 \pi^{2}} y^{2} T^{4} S_{2}^{2}(y), \\
		h=&e+\frac{P}{n}=y T \frac{S_{3}^{1}(y)}{S_{2}^{1}(y)}\label{enthalpy}.
	\end{align}
	The particle number conservation ($\partial_{\mu}N^{\mu} =0$) and 
	contraction of energy-momentum conservation with $u_{\mu}$ and 
	$\Delta_{\mu\nu}$ ($u_{\mu}\partial_{\nu}T^{\mu\nu}=0$, 
	$\Delta^{\mu}_{\nu}\partial_{\alpha}T^{\nu\alpha}=0$)  leads to the 
	continuity equation, equation of energy and equation of motion 
	respectively as follows
	\begin{align}
		&Dn = -n\partial_{\mu}u^{\mu},\\
		&De = -\frac{P}{n}\partial_{\mu}u^{\mu}\\\label{80}
		&	Du^{\mu} = \frac{1}{nh}\nabla^{\mu}P.
	\end{align}
	In the hydrodynamic regime slightly away from equilibrium, the 
	relativistic Gibbs-Duhem relation is given as
	\begin{equation}
		\partial_{\nu}P = nT\partial_{\nu}\left(\frac{\mu}{T}\right)+nhT^{-1}\left(\partial_{\nu}T\right),
	\end{equation}
	which on contraction with $u^{\mu}$ can be rewritten as
	\begin{equation}\label{82}
		Dh = TD\left(\frac{\mu}{T}\right) + hT^{-1}\left(DT\right).
	\end{equation}
	Eq.\eqref{80} and \eqref{82} can be expanded in terms of derivative of 
	temperature and chemical potential over temperature as
	\begin{align}
		&\left(\frac{\partial e}{\partial T}\right)_{\mu / T} D T+\left(\frac{\partial e}{\partial(\mu / T)}\right)_{T} D\left(\frac{\mu}{T}\right)=-\frac{P}{n} \nabla_{\mu} u^{\mu}, \\
		&{\left[\left(\frac{\partial h}{\partial T}\right)_{\mu / T}-h T^{-1}\right] D T+\left[\left(\frac{\partial h}{\partial(\mu / T)}\right)_{T}-T\right] D\left(\frac{\mu}{T}\right)=0},\label{86}
	\end{align}
	and on further using Eq.\eqref{energy_density} and \eqref{enthalpy}, we 
	can derive the following quantities
	\begin{align}
		\left(\frac{\partial h}{\partial T}\right)_{\mu / T}=&y\left[5 \frac{S_{3}^{1}}{S_{2}^{1}}+y \frac{S_{2}^{0}}{S_{2}^{1}}-y \frac{S_{3}^{1} S_{3}^{0}}{\left(S_{2}^{1}\right)^{2}}\right], \\
		\left(\frac{\partial e}{\partial T}\right)_{\mu / T}=&4 y \frac{S_{3}^{1}}{S_{2}^{1}}+y \frac{S_{2}^{2} S_{3}^{0}}{\left(S_{2}^{1}\right)^{2}}-\frac{S_{2}^{2}}{S_{2}^{1}}+y^{2}\left[\frac{S_{2}^{0}}{S_{2}^{1}}-\frac{S_{3}^{1} S_{3}^{0}}{\left(S_{2}^{1}\right)^{2}}\right], \\
		\left(\frac{\partial h}{\partial(\mu / T)}\right)_{T}=&T y\left[\frac{S_{3}^{0}}{S_{2}^{1}}-\frac{S_{3}^{1} S_{2}^{0}}{\left(S_{2}^{1}\right)^{2}}\right], \\
		\left(\frac{\partial e}{\partial(\mu / T)}\right)_{T}=&-T\left[1-\frac{S_{2}^{2} S_{2}^{0}}{\left(S_{2}^{1}\right)^{2}}\right]+T y\left[\frac{S_{3}^{0}}{S_{2}^{1}}-\frac{S_{3}^{1} S_{2}^{0}}{\left(S_{2}^{1}\right)^{2}}\right].\label{87}
	\end{align}
	We can solve for $DT$ and $D\left(\frac{\mu}{T}\right)$ using 
	the above equations and we get
	\begin{align}\label{dT}
		\frac{1}{T}DT =& \left(1-\lambda'\right)\nabla_{\mu}u^{\mu},\\
		TD\left(\frac{\mu}{T}\right) =& \left[\left(\lambda''-1\right)h-\lambda'''T\right]\nabla_{\mu}u^{\mu},\label{89}
	\end{align}
	where,
	\begin{align}
		\lambda^{\prime}=&\frac{\left(S_{2}^{0} / S_{2}^{1}\right)^{2}-\left(S_{3}^{0} / S_{2}^{1}\right)^{2}+4 y^{-1} S_{2}^{0} S_{3}^{1} /\left(S_{2}^{1}\right)^{2}+y^{-1}\left(S_{3}^{0} / S_{2}^{1}\right)}{\left(S_{2}^{0} / S_{2}^{1}\right)^{2}-\left(S_{3}^{0} / S_{2}^{1}\right)^{2}+3 y^{-1} S_{2}^{0} S_{3}^{1} /\left(S_{2}^{1}\right)^{2}+2 y^{-1}\left(S_{3}^{0} / S_{2}^{1}\right)-y^{-2}}, \\
		\lambda^{\prime \prime}=&1+\frac{y^{-2}}{\left(S_{2}^{0} / S_{2}^{1}\right)^{2}-\left(S_{3}^{0} / S_{2}^{1}\right)^{2}+3 y^{-1} S_{2}^{0} S_{3}^{1} /\left(S_{2}^{1}\right)^{2}+2 y^{-1}\left(S_{3}^{0} / S_{2}^{1}\right)-y^{-2}}, \\
		\lambda^{\prime \prime \prime}=&\frac{\left(S_{2}^{0} / S_{2}^{1}\right)+5 y^{-1}\left(S_{3}^{1} / S_{2}^{1}\right)-S_{3}^{0} S_{3}^{1} /\left(S_{2}^{1}\right)^{2}}{\left(S_{2}^{0} / S_{2}^{1}\right)^{2}-\left(S_{3}^{0} / S_{2}^{1}\right)^{2}+3 y^{-1} S_{2}^{0} S_{3}^{1} /\left(S_{2}^{1}\right)^{2}+2 y^{-1}\left(S_{3}^{0} / S_{2}^{1}\right)-y^{-2}} .
	\end{align}
	Now, replacing the time derivative using Eq.\eqref{dT} and \eqref{89} 
	in Eq.\eqref{50}, we get
	\begin{align}
		p^{\mu}\partial_{\mu}f_{eq} = \frac{1}{T}f_{eq}(1-f_{eq})\Big[\left(u\cdot p\right)^2\left(1-\lambda'\right)\nabla_{\mu}u^{\mu}+&\left(\frac{u\cdot p}{T}\right)p^{\mu}\nabla_{\mu}T+\\\nonumber
		\left(u\cdot p\right)\left\{\left(\lambda''-1\right)h-\lambda'''T\right\}\nabla_{\mu}u^{\mu}
		&+Tp^{\mu}\nabla_{\mu}\left(\frac{\mu}{T}\right)\\\nonumber
		-p_{\nu}\left\{\left(u\cdot p\right)Du^{\nu}+p^{\mu}\nabla_{\mu}u^{\nu}\right\}\Big].
	\end{align}
	Employ the equation of motion
	with relativistic Gibbs-Duhem relation, we finally get,
	\begin{eqnarray}
		p^{\mu}\partial_{\mu}f_{eq}=	\frac{f_{eq}\left(1-f_{eq}\right)}{T}\Bigg[\Big\{\left(u\cdot p\right)^2\left(\frac{4}{3}-\lambda'\right)-\frac{1}{3}m^2+&\\\nonumber
		\left(u\cdot p\right)\big(\left(\lambda''-1\right)h-\lambda'''T\big)\Big\}\nabla_{\mu}u^{\mu}-p^{\mu}p^{\nu}\left(\nabla_{\mu}u_{\nu}-\frac{1}{3}\Delta_{\mu\nu}\nabla_{\sigma}u^{\sigma}\right)&+\\\nonumber
		\left(1-\frac{u\cdot p}{h}\right)Tp^{\mu}\nabla_{\mu}\left(\frac{\mu}{T}\right)\Bigg].
	\end{eqnarray}
	\section{Thermodynamic integrals}\label{append_c}	
	We define the thermodynamic functions,
	\begin{equation}
		J_{nq}^{(m)\pm}=\sum_{f}\frac{1}{(2q+1)!!}\int dP (u\cdot p)^{n-2q-m}\left(\Delta_{\alpha\beta}p^{\alpha}p^{\beta}\right)^{q}\left(f_{eq,f}\tilde{f}_{eq,f}\pm \bar{f}_{eq,f}\tilde{\bar{f}}_{eq,f}\right),
	\end{equation}
	
	where $dP = g_fd^{3}p/\left[(2\pi)^3\varepsilon_f\right], \tilde{f}_{eq}=1-\alpha f_{eq}.$ Similarly, we define
	\begin{equation}
		{Y_{nq}^{(m)\pm}=\sum_{f}\frac{1}{(2q+1)!!}\int dP \frac{(u\cdot p)^{n-2q-m}}{\left[1+\left(\frac{q_fB\tau_f}{u\cdot p}\right)^2\right]}\left(\Delta_{\alpha\beta}p^{\alpha}p^{\beta}\right)^{q}\left(f_{eq,f}\tilde{f}_{eq,f}\pm \bar{f}_{eq,f}\tilde{\bar{f}}_{eq,f}\right),}
	\end{equation}
	and
	\begin{equation}
		{Z_{nq}^{(m)\pm}=\sum_{f}\frac{1}{(2q+1)!!}\int dP \frac{(u\cdot p)^{n-2q-m}}{\left[1+\left(\frac{2q_fB\tau_f}{u\cdot p}\right)^2\right]}\left(\Delta_{\alpha\beta}p^{\alpha}p^{\beta}\right)^{q}\left(f_{eq,f}\tilde{f}_{eq,f}\pm \bar{f}_{eq,f}\tilde{\bar{f}}_{eq,f}\right).}
	\end{equation}

\end{document}